\documentclass[useAMS,usenatbib,color]{mn2e}
\usepackage{natbib}
\usepackage{amssymb,amsmath,amsfonts}
\usepackage{epsfig}
\def\reff@jnl#1{{\rm#1\/}}

\def\aj{\reff@jnl{AJ}}                  % Astronomical Journal
\def\araa{\reff@jnl{ARA\&A}}            % Annual Review of Astron and Astrophys
\def\apj{\reff@jnl{ApJ}}                        % Astrophysical Journal
\def\apjl{\reff@jnl{ApJ}}               % Astrophysical Journal, Letters
\def\apjs{\reff@jnl{ApJS}}              % Astrophysical Journal, Supplement
\def\ao{\reff@jnl{Appl.Optics}}         % Applied Optics
\def\apss{\reff@jnl{Ap\&SS}}            % Astrophysics and Space Science
\def\aap{\reff@jnl{A\&A}}               % Astronomy and Astrophysics
\def\aapr{\reff@jnl{A\&A~Rev.}}         % Astronomy and Astrophysics Reviews
\def\aaps{\reff@jnl{A\&AS}}             % Astronomy and Astrophysics, Supplement
\def\azh{\reff@jnl{AZh}}                        % Astronomicheskii Zhurnal
\def\baas{\reff@jnl{BAAS}}              % Bulletin of the AAS
\def\jrasc{\reff@jnl{JRASC}}            % Journal of the RAS of Canada
\def\memras{\reff@jnl{MmRAS}}           % Memoirs of the RAS
\def\mnras{\reff@jnl{MNRAS}}            % Monthly Notices of the RAS
\def\pra{\reff@jnl{Phys.Rev.A}}         % Physical Review A: General Physics
\def\prb{\reff@jnl{Phys.Rev.B}}         % Physical Review B: Solid State
\def\prc{\reff@jnl{Phys.Rev.C}}         % Physical Review C
\def\prd{\reff@jnl{Phys.Rev.D}}         % Physical Review D
\def\prl{\reff@jnl{Phys.Rev.Lett}}      % Physical Review Letters
\def\pasp{\reff@jnl{PASP}}              % Publications of the ASP
\def\pasj{\reff@jnl{PASJ}}              % Publications of the ASJ
\def\qjras{\reff@jnl{QJRAS}}            % Quarterly Journal of the RAS
\def\skytel{\reff@jnl{S\&T}}            % Sky and Telescope
\def\solphys{\reff@jnl{Solar~Phys.}}    % Solar Physics
\def\sovast{\reff@jnl{Soviet~Ast.}}     % Soviet Astronomy
\def\ssr{\reff@jnl{Space~Sci.Rev.}}     % Space Science Reviews
\def\zap{\reff@jnl{ZAp}}                        % Zeitschrift fuer Astrophysik
\def\nat{\reff@jnl{Nature}}             % Nature 
\def\physrep{\reff@jnl{Phys.~Rep.}}    % Physics Reports

\title[Cross-correlating 21-cm, IR, and SZ Backgrounds]{
Cross-correlation studies as a probe of reionization physics}
\label{firstpage}
\author[Slosar, Cooray \& Silk] {An\v{z}e Slosar$^{1,2}$, Asantha Cooray$^3$,
  Joseph I. Silk$^1$\\
  $^1$Oxford Astrophysics, Denys Wilkinson Building, Keble Road, OX13RH,
  Oxford, United Kingdom \\
  $^2$Faculty of Mathematics and Physics, University of Ljubljana,
  Slovenia \\
  $^3$Center for Cosmology, Department of Physics and Astronomy, University of California, Irvine, CA 9269, USA}

  \date{Accepted ---; received ---; in original form \today}
\pagerange{\pageref{firstpage}--\pageref{lastpage}}
\pubyear{2002}

\newcommand {\rmd}{{\rm d}}

\newcommand {\Mmin}{M_{\rm min}}
\newcommand {\Mprog}{M_{\rm prog}}
\newcommand {\sigmat}{\sigma_{\rm T}}
\newcommand {\sn}{{\rm sn}}

\newcommand{\mpch}{{h^{-1}\ \rm Mpc}}

\newcommand{\vect}[1]{\mathbf{#1}}
\newcommand{\fstar}{f_\star}
\newcommand{\fb}{f_{\rm b}}

\begin{document}
\maketitle

\begin{abstract} 
  The process of reionization is now believed to have proceeded in an
  orchestrated manner beginning with UV photons emitted by high
  redshift galaxies containing a large fraction of Population III
  stars carving out ionised regions around them. The physics during
  this era can be studied with a combination of redshifted 21-cm
  spin-flip transition tracing neutral hydrogen gas, IR emission from
  massive primordial stars that trace the global star-formation rate
  during reionization, and the imprint of hot-electrons in first
  supernovae remnants Compton-cooling off of cosmic microwave
  background (CMB) radiation through the Sunyaev-Zel'dovich effect.
  While these individual effects and their observable signatures have
  been advocated as probes of reionization history, here we show how
  cross-correlation studies between these signals can be used to
  further understand physics during reionization.  Cross-correlation
  studies are advantageous since the measurable statistics do not
  suffer in the same manner from foregrounds and systematic effects as
  is the case of auto-correlation function measurements.  We discuss
  the prospects for detecting various cross-correlation statistics
  using present and next generation experiments and the information
  related to reionization captured by them.
\end{abstract}

\begin{keywords}
  cosmology: theory --- large scale structure --- infrared: general --- stars: formation --- cosmology: observations --- diffuse radiation
\end{keywords}

\section{Introduction}

The 21-cm spin-flip transition of neutral Hydrogen, either in the form
of an absorption or an emission relative to Cosmic Microwave
Background (CMB) blackbody spectrum, provides one of the best ways to
study the intergalactic medium during and prior to reionization
\citep{1990MNRAS.247..510S,Tozzi:1999zh,Madau:1996cs,Furlanetto:2003nf,Santos:2006fp}.
With frequency selection for observations, the 21-cm line, in
principle, provides three-dimensional tomography of the reionization
era as well as a probe to the dark ages where no luminous sources are
present.  The exact physics associated with the reionization process
is still largely unknown, though it is strongly believed that UV
photons from first luminous sources are responsible for it
\citep{Barkana:2000fd,2006PhR...433..181F}. These UV photons create
bubbles
\citep{Wyithe:2003rr,Cen:2003ry,Cen:2002zc,Haiman:2003ea,Mackey:2002yn,Santos:2003jb,
  Yoshida:2003rw,Yoshida:2003ab,2006astro.ph..4177Z} of ionised
material around them, although it is still unclear whether densest or
least-dense regions were ionised first
\citep{2005MNRAS.363.1031F,2006astro.ph..3438C}. Direct detection of
auto-correlation spectra of this signal might prove to be difficult,
due to very strong foregrounds, although various cleaning techniques
have been proposed
\citep{Zaldarriaga:2003du,Santos:2004ju,Morales:2005qk}

The same reionization redshifts can also be probed at near-IR
wavelengths since the intensity of the cosmic near-infrared background
(IRB) is a measure of the total light emitted by stars and galaxies in
the Universe. The possibility that there is a high-redshift component
to the IRB comes from the fact that the absolute background estimated
by space-based experiments, such as the Diffuse Infrared Background
Experiment (DIRBE; \cite{2001ARA&A..39..249H}) and the Infra-Red Telescope in
Space (IRTS; \cite{Matsumoto:2004dx}), is significantly larger that the
background accounted for by resolved sources so far: only 13.5 $\pm$
4.2 nW m$^{-2}$ sr$^{-1}$ is resolved to point sources at 1.25 $\mu$m
\citep{Cambresy:2001ei}, while current direct measurements range from
25-70 nW m$^{-2}$ sr$^{-1}$ (see \cite{Kashlinsky:2005mn}
for a recent review).

Primordial galaxies at redshifts 8 and higher, especially those
involving Population III stars, are generally invoked to explain the
missing IR flux between 1 $\mu$m and 2 $\mu$m, with most of the
intensity associated with redshifted Lyman- $\alpha$ emission during
reionization \citep{Santos:2002hd,Salvaterra:2002rg,Cooray:2004gx,
  Fernandez:2005gx}.  While models of high-redshift Pop III
populations can explain the ``missing'' IRB, these models run into
several difficulties if such sources were to account for all of the
missing IR intensity. These include the high efficiency required to
convert baryons to stars in first galaxies \citep[see
e.g.][]{Madau:2005wv} and limits from deep IR imaging data that
suggest a lack of a large population of high-redshift dropouts
\citep{Salvaterra:2005ga}.  Still, one does expect some contribution
to the IRB from sources that reionized the Universe, though the exact
intensity of the IRB from such sources is yet unknown both
theoretically and observationally.  

As pointed out in
\cite{Cooray:2003yf} \citep[see also][]{Kashlinsky:2004xx}, if a
high-redshift population contributes significantly to the IRB, then
these sources are expected to leave a distinct signal in the
anisotropy fluctuations of the near-IR intensity, when compared to the
anisotropy spectrum associated with low-redshift sources. Such a study
has been attempted with anisotropy measurements using Spitzer data
with preliminary indications for an excess anisotropy in the
background at arcminute angular scales and below
\citep{Kashlinsky:2005di}.  There are, however, large uncertainties on
the exact fraction of IRB intensity from redshifts during reionization
\citep{Sullivan:2006vr,Salvaterra:2005rd}, especially given the suggestions that these
excess IRB fluctuations can be partly explained with expected clustering from faint unresolved galaxies in optical
images but not present in IR images \citep{Cooray:2006dc}.   Here, we suggest that
an approach to establish the presence of IRB anisotropies from
the era of reionization, such as those due to first-stars, will be to
consider a cross-correlation of the IRB against brightness temperature
fluctuations in the 21-cm background.

Finally, the small-scale cosmic microwave background (CMB)
anisotropies are also expected to contain signatures from the
reionization epoch
\citep{Santos:2003jb,2005ApJ...630..657Z,2006NewAR..50...84M,2005astro.ph.12263M,Alvarez:2005sa}.
Uniquely identifying the fractional anisotropy contribution from
reionization era is challenging with CMB data alone since a large
number of sources contribute to the anisotropies of CMB at arcminute
angular scales and below. These include thermal Sunyaev-Zel'dovich
(SZ) effect from low-redshift galaxy clusters, the Ostriker-Vishniac
effect associated with bulk motions of the electron density field, and
gravitational lensing modifications to CMB anisotropies. In terms of
reionization contributions, one expects an anisotropy when electrons
in the surrounding medium of first supernovae explosions Compton-cool
against the CMB \citep{Oh:2003sa}; the modification to CMB is similar
to the y-distortion related to SZ scattering in clusters.  Such a
signal need not be associated with supernovae alone, as X-rays from
first quasars, supernovae, and star clusters are also expected to heat
the intergalactic medium (IGM) at high redshifts
\citep{Oh:2000zx,Venkatesan:2001cd}.  Such a high-redshift
contribution could also help explain the excess CMB anisotropy at
arcminute angular scales, as seen by CBI \citep{2003ApJ...591..540M},
BIMA \citep{2002ApJ...581...86D,Dawson:2006qd}, and ACBAR
\citep{2004ApJ...600...32K,Kuo:2006ya} experiments.

If a high-redshift SZ contribution exists, an interesting possibility
to identify it involves the cross-correlation between small-scale CMB
anisotropy maps and the 21-cm signal.  If the IGM is heated to a
temperature above the CMB, as due to X-ray heating, then the 21-cm
signal will be detectable as an emission.  While SZ originates from
the ionized electrons and the 21-cm signal is related to the neutral Hydrogen
distribution, with partial reionization, one expects the two signals
to spatially anti-correlate as the regions containing free-electrons
trace the same underlying density field defined by dark matter that
also leads to fluctuations in the neutral Hydrogen.  The
cross-correlation between arcminute-scale CMB and the 21-cm can be
used to measure the angular power spectrum of this (anti)-correlation
and help establish the fraction of SZ effect originating from
high-redshifts.

In this paper, we discuss several cross-correlations involving 21-cm
data and tracers of reionization discussed above.  Since 21-cm
observations can be preselected based on the choice of a frequency bin
corresponding to the redshifted line emission, one can consider the
correlation as a function of the redshift bin and use that information
to establish the redshift dependence of the high-redshift SZ signal as
well as the IRB component related to primordial galaxies.  Here, we
discuss the measurement of the proposed cross-correlations using CMB
maps from upcoming missions and maps of the IRB from wide-field
images. A cross-correlation of the 21-cm signal with galaxy surveys
has been discussed in \cite{Furlanetto:2006pg} and
\cite{Wyithe:2006vg}.

In this paper we use a simple analytic model to calculate and
understand the qualitative features of the cross-correlation. For more
accurate quantitative predictions, one should use the numerical
simulations, which have recently made significant advances
\citep{Iliev:2005sz,Mellema:2006pd, Iliev:2006sw,Kohler:2005gg,2006astro.ph..4177Z}.

This paper is organized as following. In Section~\ref{sec:method}, we
present our basic model for the auto- and cross-correlations of
various fields discussed here. In Section~\ref{sec:cross-corr-calc},
we consider a physical model of reionization and discuss our results
and observability of various statistics.  Finally, we conclude with a
summary of our results in Section~\ref{sec:disc-concl}.

\section{Method}
\label{sec:method}
\newcommand{\hvect}[1]{\mathbf{\hat{#1}}}

\begin{figure}
  \centering
\begin{tabular}{cc}
  \includegraphics[width=0.4\linewidth]{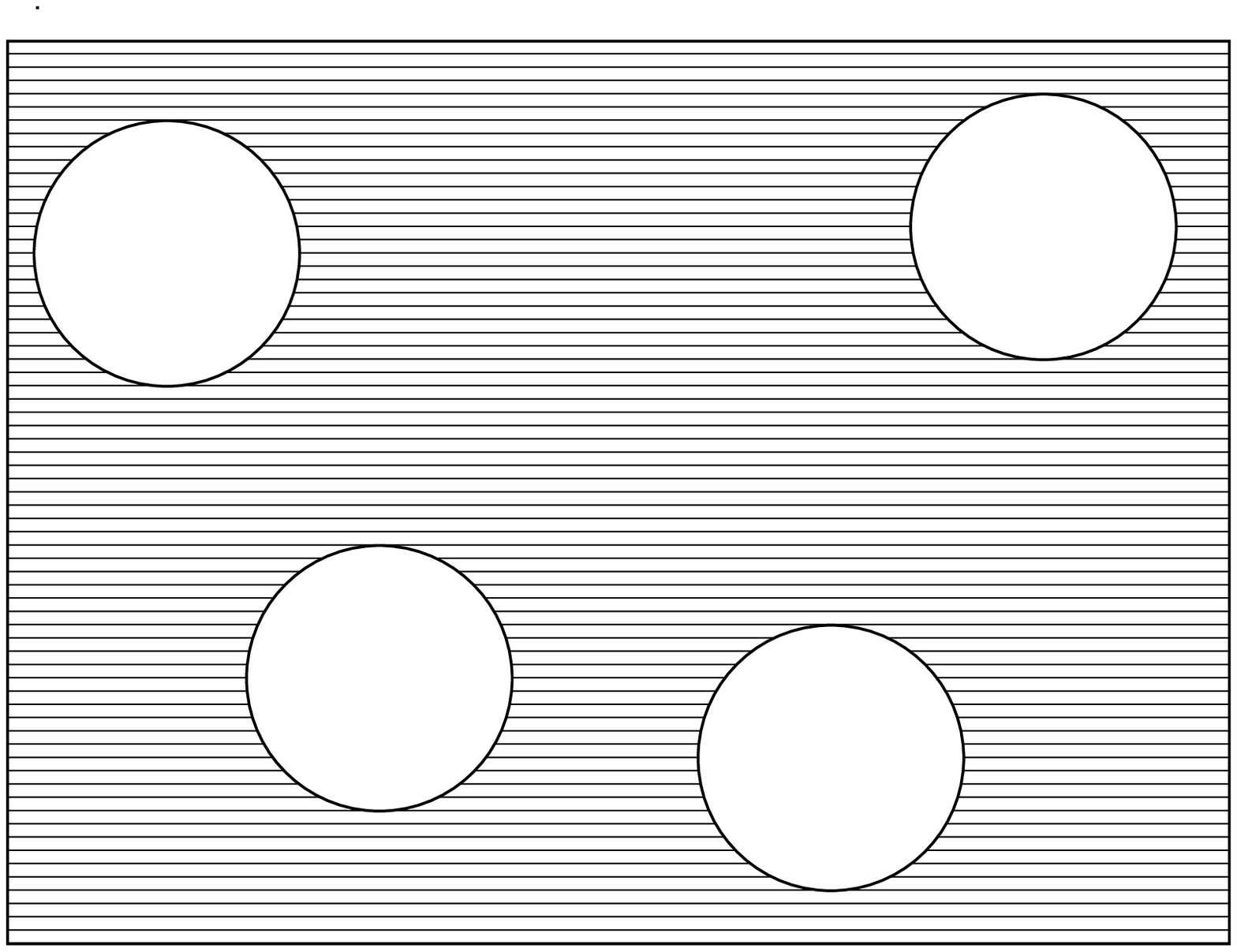}  &
  \includegraphics[width=0.4\linewidth]{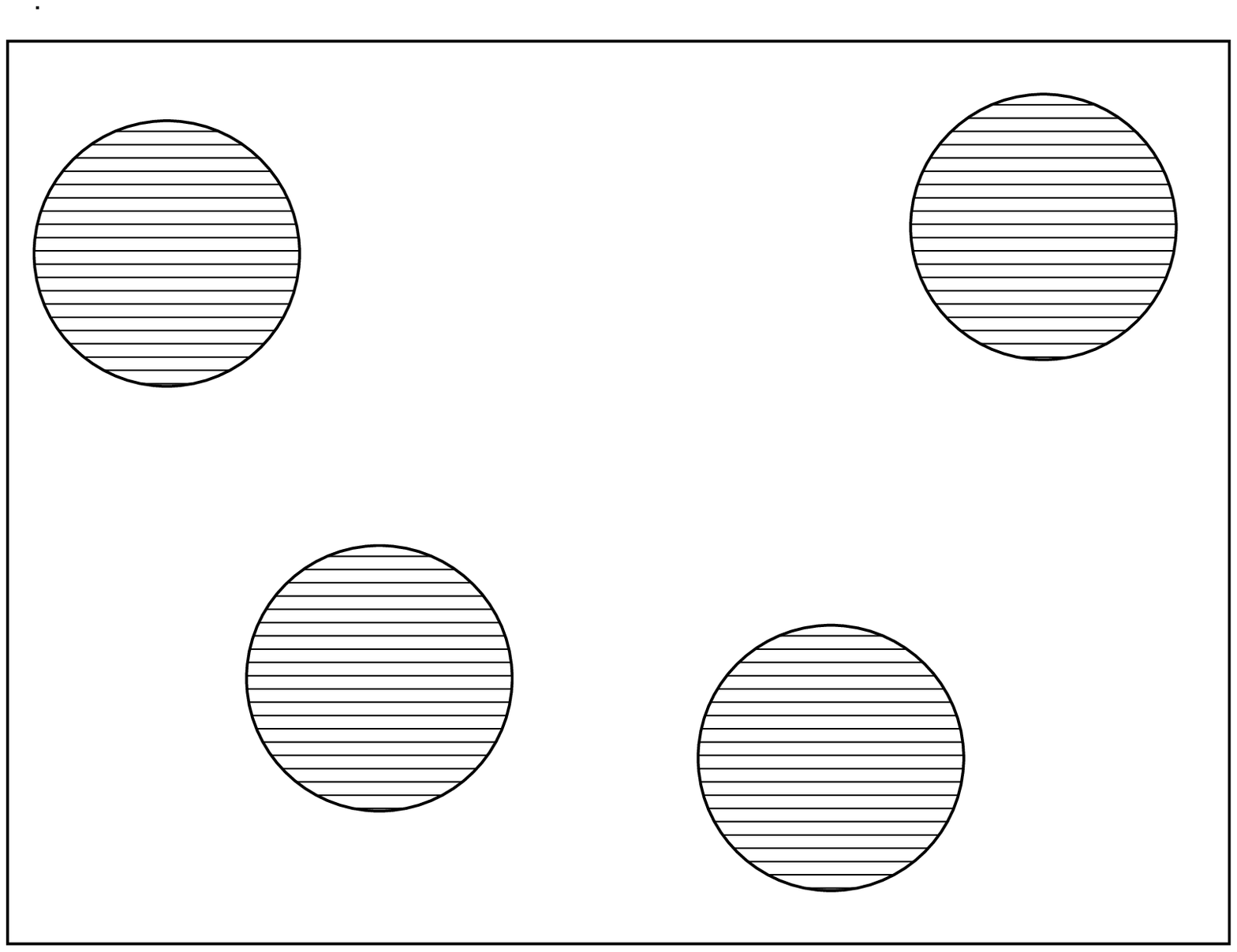}  \\
  $x_H$ & $1-x_H$ \\
  \includegraphics[width=0.4\linewidth]{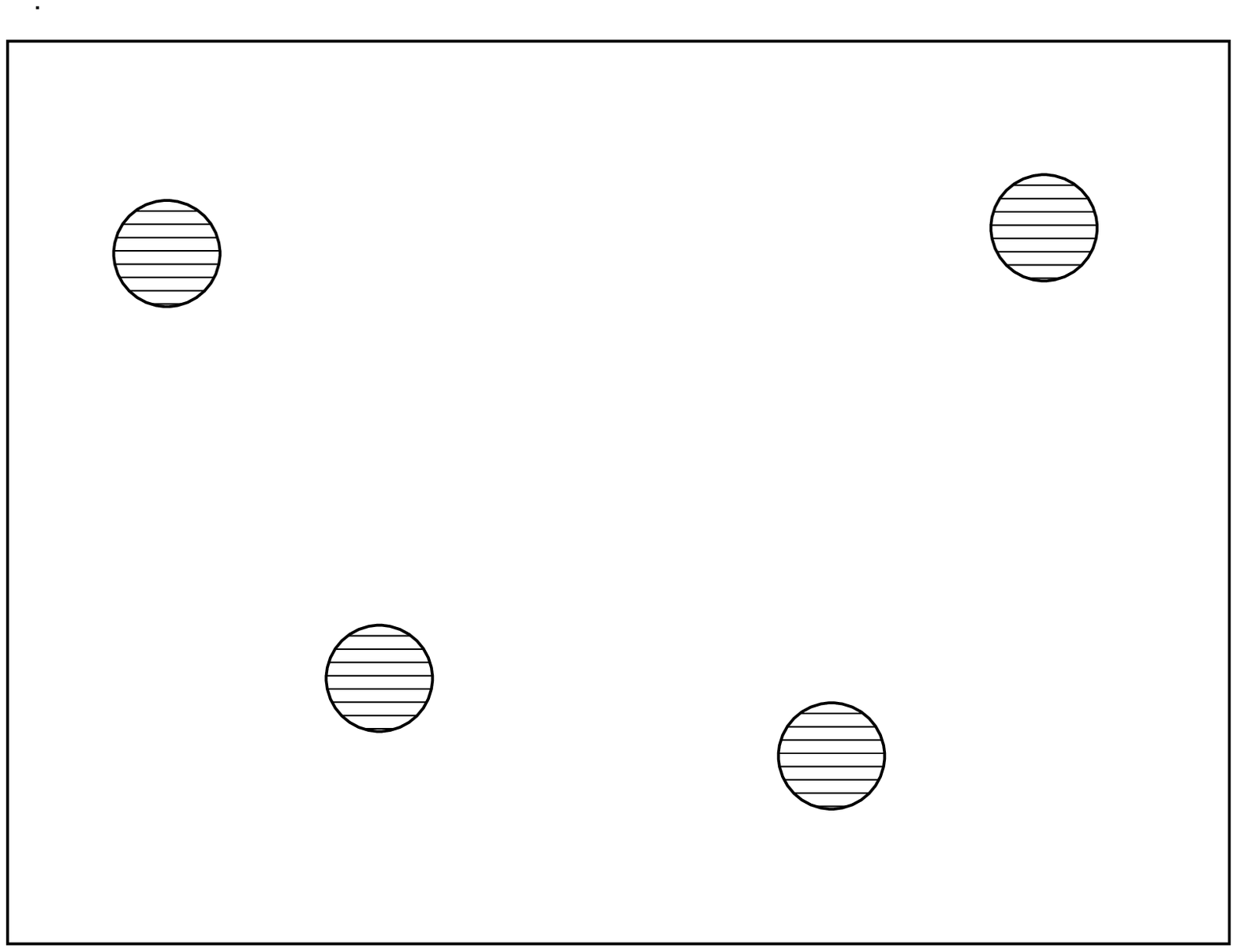} & \\
 $x_s$ & \\
\end{tabular}
  \caption{This figure shows schematically, the neutral fraction
    field $x_H$ (top left), the ``negative'' neutral fraction field
    (top right) and the sources field  $x_S$ (bottom) used in this paper. }
 \label{fig:ig123}
\end{figure}

In this Section, we  develop simple models for 21cm, high-redshift SZ, and IR
backgrounds and their cross-correlation. To describe the 21-cm background, we closely follow the
formalism developed in \cite{Zaldarriaga:2003du} and extend it to
encompass the IR and SZ fluctuations as well. The basic starting
premise is that the 21-cm fluctuations are tracing the dimensionless
brightness temperature field
\begin{equation}
\label{eq:2}
\psi(\hvect{n},r)=(1+\delta)x_H,
\end{equation}
where $\delta$ is the fraction over-density and $x_H$ is the
neutral fraction, while the IR and SZ fluctuations trace the
``sources'' field
\begin{equation}
\label{eq:6}
\phi(\hvect{n},r)=(1+\delta_s) \,.
\end{equation}
If we assume that during reionization, the early sources form bubbles
of nearly completely ionised gas so that $x_H$ is strongly bimodal and
that this gas is everywhere hot enough to produce SZ fluctuations and
that it is filled with primordial sources emitting in (observed-frame)
infrared light, then it is reasonable to expect that
$\phi(\hvect{n},r) \propto (1+\delta)x_S$ with $x_S=1-x_H$. Here we
take a marginally more realistic model, by assuming that sources occupy
just some central fraction of the reionization halos as illustrated
in the Figure \ref{fig:ig123}. To what extend this is
a plausible approximation remains to be seen, but it is a useful
starting point for the cross-correlation study we propose here.

We will use superscripts $n$, $i$, $y$ to describe the 21-cm spin flip
transition, $i$ for the IR signal and $y$ for the SZ signal from the
epoch of reionization. For example, $C_\ell^{ni}$ denotes the angular
cross-correlation power spectrum between the 21-cm and the IR signals.
We will adopt units of mK for th 21-cm signal, nW/m$^2$/sr for the IR
signal and the observed $\mu$K decrement in the Rayleigh-Jeans region
for the Comptonization parameter ($y$). The latter is given by $\Delta
T = -2 T_{\rm CMB} y$.

Through the paper we use a standard flat cosmology with $\Omega_{\rm
 b}=0.05$, $\Omega_{m}=0.3$, $\Omega_\Lambda=0.7$, $\sigma_8=0.8$ and
$n_{\rm s}=1$, where the symbols have their standard meaning.

\subsection{21cm background and power spectrum}

In the limit of no redshift distortions and spin temperature of
hydrogen being much larger than the CMB temperature, the observed
brightness temperature on the sky is given by
\citep{Zaldarriaga:2003du}
\begin{equation}
  \label{eq:1}
  T(\hvect{n}) = T_0(r_0) \int \rmd r W(r) \psi (\hvect{n},r),
\end{equation}
were $W(r)$ is the window function describing the instrumental
bandwidth around some target redshift $r_0$, normalized such that $\int
\rmd r W(r) = 1$ and
\begin{equation}
  T_0(r) = 23{\rm mK}\left(\frac{\omega_b}{0.02}
  \right)\left(\frac{\omega_m}{0.15}\right)^{-1/2} \left(\frac{1+z}{10} \right),
\end{equation}
with $\psi$ being defined in equation \eqref{eq:2}

If we expand $\psi$ in Fourier series
\begin{equation}
  \label{eq:3}
  \psi(\vect{x}) = \int \frac{\rmd^3k}{(2\pi)^3} \psi_k(\vect{k})
  e^{i\vect{k}\cdot\vect{r}}
\end{equation}
and then use the spherical harmonic decomposition of the observed temperature
\begin{eqnarray}
 && a_{\ell m} = 4 \pi i^\ell \int \frac{\rmd^3k}{(2\pi)^3}
  \psi_k(\vect{k}) \alpha_\ell(k,\nu) Y^{\star}_{\ell
    m}(\vect{k}) \nonumber \\
&&\alpha_\ell(k,\nu)= T_0(r_0) \int \rmd r W(r) j_\ell(kr),
\end{eqnarray}
we can define the angular power spectrum 
\begin{equation}
  \label{eq:5}
  \left<a_{\ell m} a^\star_{\ell' m'}  \right>=\delta_{\ell
    \ell'}\delta_{m m'} C_\ell^{nn}, 
\end{equation}
so that
\begin{equation}
\label{eq:nn}
  C_\ell^{nn} = 4 \pi \int \frac{\rmd k}{k} \Delta_{\psi\psi}^2(k)
  \alpha_\ell^2(k), 
\end{equation}
where
\begin{equation}
  \Delta^2_{\psi\psi}(k) = \frac{k^3}{2\pi^2} P_{\psi\psi}(k) =  \frac{k^3}{2\pi^2}
  \delta^{D}(\vect{k}+\vect{k'})\left< \psi_k(\vect{k})\psi_k(\vect{k}')\right> 
\end{equation}

\subsection{IR background and power spectrum}

We now repeat the exercise for the IR flux from the first stars. The
specific intensity is given by
\begin{equation}
  \label{eq:21}
  I_\nu(\hvect{n}) =  \int\rmd r  \frac{\rmd I_\nu}{\rmd r} \phi (\hvect{n},r),
\end{equation}
with
\begin{equation}
  \frac{\rmd I_\nu}{\rmd{r}} = \frac{1}{\bar{\phi}(r)}\frac{\rmd
    V}{\rmd r}\Psi(r) 
t_\star \frac{l_\star(\nu(1+z))(1+z)}{4\pi D_l^2},
\end{equation}
where $\bar{\phi}(r)$ is the average value of $\phi$ (defined in
equation \eqref{eq:6}) at a given distance from the observer. Other
quantities take care of producing the correct mean intensity: $\Psi$
is the star-formation rate per unit volume at a given redshift,
$t_\star \sim 5 \times 10^6{\rm yr}$, $l_\star$ is the mean spectrum per unit
mass of a Pop III star and $D_l$ is the luminosity distance.  We take
the spectrum of first stars from \cite{Santos:2002hd} assuming unity
escape fraction. Assuming a different escape fraction would change our
results, but quantitatively the would remain the same. 

\begin{figure}
  \centering
  \includegraphics[width=1.0\linewidth]{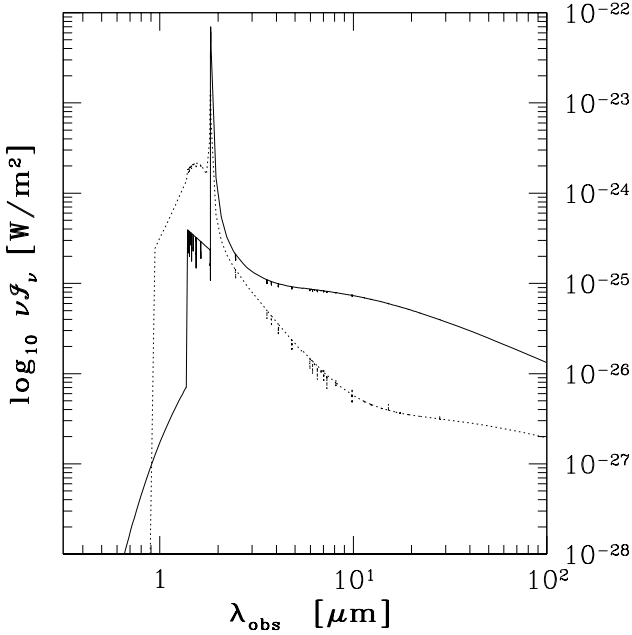}  
  \caption{The observed spectrum of 100 solar mass
  primordial Pop-III star at redshift of $z=15$ as a function of the
  observed wavelength. The solid line is for unity escape fraction and the
  dotted line is for zero escape fraction - see \citep{Santos:2002hd}
  for details.}
 \label{fig:pop3}
\end{figure}

In Figure~\ref{fig:pop3}, we plot intensity spectra for a 100 $M_\odot$ mass star at
redshift of $z=15$, but observed today.  Note the pronounced (even with a logarithmic
scale!) Lyman-alpha peak (due to collisionally excited atoms, see
\cite{Santos:2002hd}) at around few microns in wavelength. At lower wavelengths,
this emission dominates the total intensity signal from such a star. This means that if our observing wavelength
is such that the integral includes a contribution from the peak, then
the resulting intensity power spectrum would be dominated by this emission. 

If we expand $\phi$ in Fourier series and use the spherical harmonic
decomposition in exact analogy with equations
\eqref{eq:3}--\eqref{eq:nn} we can define the angular power spectrum
given by
\begin{equation}
\label{eq:ii}
  C_\ell^{ii} = 4 \pi \int \frac{\rmd k}{k} \Delta_{\phi\phi}^2(k)
  \beta_\ell^2(k), 
\end{equation}
where
\begin{equation}
\beta_\ell(k,\nu)= \int \rmd r \frac{\rmd I_\nu(r)}{\rmd r} j_\ell(kr)
\end{equation}
and
\begin{equation}
  \Delta^2_{\phi\phi}(k) = \frac{k^3}{2\pi^2} P_{\phi\phi}(k) =  \frac{k^3}{2\pi^2}
  \delta^{D}(\vect{k}+\vect{k'})\left< \phi_k(\vect{k})\phi_k(\vect{k}')\right> \\
  \label{eq:24}
\end{equation}

\subsection{High-redshift SZ component}

The high-redshift SZ component is handled in a very similar manner.
We write the total Comptonization parameter as 
\begin{equation}
  \label{eq:31}
  y(\hvect{n}) =  \int\rmd r  \frac{\rmd y}{\rmd r} \phi (\hvect{n},r) \, .
\end{equation}
To calculate the Comptonization parameter, we model each supernovae as
a spherical ball of radius $R$. The mean Comptonization parameter per
SN remnant is $y = k_{\rm b} T_{\rm e}/(m_e c^2) \sigmat n_e 4R/3$
($\sigmat$ is the Thomson scattering cross-section, $m_e$ is the
electron mass) and each remnant contributes over $\pi (R/D_A)^2$ per
steradian to mean $\bar{y}$.  Assuming that SN energy $E_{\sn} \sim
10^{44} J$ is evenly distributed between electrons ($E_{\sn}=N_e
k_{\rm b} T_e$) one arrives at
\begin{equation}
  \frac{\rmd y}{\rmd r} = \frac{1}{\bar{\phi}(r)}\frac{\rmd V}{\rmd r}  \eta_{\sn}
  \frac{\Psi}{\Mprog} t_{\sn} \left(\frac{E_{\sn} }{m_e c^2} \right)
  \left(\frac{\sigmat}{D_A^2}\right),
\end{equation}
where $\Mprog\sim50 M_\odot$ is a typical progenitor mass, and
$\eta_{\sn}\sim0.5-1$ takes into account the fact that not all stars
end with a supernovae and that not all energy is deposited into hot
electrons.

We can we can now define the angular power spectrum given by
\begin{equation}
\label{eq:yy}
  C_\ell^{yy} = 4 \pi \int \frac{\rmd k}{k} \Delta_{\phi\phi}^2(k)
  \gamma_\ell^2(k), 
\end{equation}
where
\begin{equation}
\gamma_\ell(k)= \int \rmd r \frac{\rmd y(r)}{\rmd r} j_\ell(kr).
\end{equation}

\subsection{The cross-correlation power spectrum}
When cross-correlating with the 21-cm, the cross-correlation power
spectra are now defined in exactly the same manner:
\begin{eqnarray}
\label{eq:x}
  C_\ell^{ni} = 4 \pi \int \frac{\rmd k}{k} \Delta^2_{\psi\phi}(k)
  \alpha_\ell(k) \beta_\ell(k)\\
  C_\ell^{ny} = 4 \pi \int \frac{\rmd k}{k} \Delta^2_{\psi\phi}(k)
  \alpha_\ell(k) \gamma_\ell(k)\\
  C_\ell^{iy} = 4 \pi \int \frac{\rmd k}{k} \Delta^2_{\phi\phi}(k)
  \beta_\ell(k) \gamma_\ell(k)
\end{eqnarray}
where
\begin{equation}
  \Delta^2_{\psi\phi}(k) = \frac{k^3}{2\pi^2} P_{\psi\phi}(k) =  \frac{k^3}{2\pi^2}
  \delta^{D}(\vect{k}+\vect{k'})\left< \psi(\vect{k})\phi(\vect{k}')\right>  \, ,
\end{equation}
is the three-dimensional cross-correlation between 21-cm brightness
temperature and source fields. In above, indices $i$, $n$ and $y$
represent anisotropies in the IR, 21-cm brightness temperature
(neutral Hydrogen), and the CMB (due to SZ y-parameter), respectively.

\section{Cross-correlation Calculation}
\label{sec:cross-corr-calc}

To proceed, we need to produce a model of correlations of the neutral
fraction. If we assume that a) bubbles are uncorrelated and b) that
any pocket of gas is either completely ionised or completely neutral,
then a good model is \citep{Zaldarriaga:2003du}\citep[see also][]{2004ApJ...613...16F}
\newcommand{\bhx}{\bar{x}_H}
\newcommand{\shx}{\bar{x}_S}
\begin{equation}
  \left<x_H(\vect{x}) x_H(\vect{x}') \right> = \bhx^2 + (\bhx-\bhx^2)f(r/R_b),
\end{equation}
where $r=|\vect{x}-\vect{x}'|$, $R$ is typical (or effective) bubble
size, $f(x)$ goes to unity at zero and to zero at infinity.  So defined auto-correlation function has the desired
property that $ \left<x_H(\vect{x}) x_H(\vect{x}') \right> \rightarrow
\bhx^2$ as $r\rightarrow \infty$ (bubbles are uncorrelated!) and
$\left<x_H(\vect{x}) x_H(\vect{x}') \right> \rightarrow \bhx$ as
$r\rightarrow 0$ (because $x_H$ is either 1 or 0).

Similarly, the auto-correlation of the $x_S$ field satisfies
\begin{equation}
  \left<x_S(\vect{x}) x_S(\vect{x}') \right> = \shx^2 + (\shx-\shx^2)f(r/R_s),
\end{equation}
with the condition that  $\shx=(1-\bhx)(R_s/R_b)^3$ (see Figure
\ref{fig:ig123}).
To calculate the cross-correlation correctly, one would need to
calculate it given function form of $f(x)$. Here we approximate it
with one that has correct limits, i.e.
\begin{equation}
  \left<x_H(\vect{x}) x_S(\vect{x}') \right> = (\shx\bhx)\left[1-f(r/R_{bs})\right].
\end{equation}
In this calculation we have simply assumed a constant bubble size of
$R_b=4$Mpc/h with sources size $R_s=2$Mpc/h and the cross-correlation
size $R_{bs}=2.25$Mpc/h.

We can now calculate auto and cross correlators of $\psi$ and $\phi$.
These are given by 

\begin{multline}
  \left< \psi(\vect{x}) \psi(\vect{x}')\right> -
  \left<\psi\right>^2 = 
\left[ \bhx^2 + (\bhx-\bhx^2)f(r/R_b) \right] \zeta(r) +\\
(\bhx-\bhx^2)f(r/R_b) + \eta_H(r)[2\bhx+\eta_H(r)]  
\end{multline}
\begin{multline}
  \left< \phi(\vect{x}) \phi(\vect{x}')\right> -
  \left<\phi\right>^2 = 
\left[ \shx^2 + (\shx-\shx^2)f(r/R_s) \right] \zeta(r) +\\
(\shx-\shx^2)f(r/R_s) + \eta_S(r)[2\shx+\eta_S(r)]  
\end{multline}
\begin{multline}
  \left< \psi(\vect{x}) \phi(\vect{x}')\right> -
  \left<\phi\right> \left<\psi\right> =
\left[ (\bhx\shx)(1-f(r/R_{bs}) \right] \zeta(r) -\\
(\bhx\shx)f(r/R_{bs}) + \bhx\eta_S(r)+\shx\eta_H(r)+\eta_H\eta_S(r)
\end{multline}
where $\eta_{H,S}$ is the cross-correlation between neutral fraction and
density $\eta_{H,S}(r)=\left<\delta(\vect{x})x_{H,S}(\vect{x}')\right>$ and
$\zeta$ is the density contrast correlation function
$\zeta(r)=\left<\delta(\vect{x})\delta(\vect{x}')\right>$.

Assuming $\eta=0$, we can now Fourier transform these to obtain the
relevant power spectra:
\begin{eqnarray}
 P_{\psi\psi}(k)= \bhx^2 P(k) +
 (\bhx-\bhx^2)(P_{f\delta}(k)+P_f(k))\\
 P_{\phi\phi}(k)= (\shx)^2 P(k) +
 (\shx-\shx^2)(P_{f\delta}(k)+P_f(k))\\
 P_{\psi\phi}(k)=  (\bhx\shx) (P(k)- P_{f\delta}(k)-P_f(k)),\\
\end{eqnarray}
where $P_f(k)$ is the Fourier transform of $f(r)$, using the
appropriate radius $R_b$, $R_s$ or $R_{bs}$ and $P_{f\delta}$ is
the Fourier transform of $f\xi$.

Finally, we take the highly biased nature of primordial sources into
account heuristically by using a biasing factor. The mean bias is
given by
\begin{equation}
  \bar{b} = \frac{\int_{M_{\rm min}}^{\infty} \rmd M  \rmd N / \rmd
    M M b(M,z)}{\int_{M_{\rm min}}^{\infty} \rmd M  \rmd N / \rmd
    M M}
\end{equation}
We multiply the linear power spectrum by $\bar{b}^2$ for IR and SZ and
their cross-correlation spectra and by the $\bar{b}(z)$ for the cross
power spectra of 21cm with either IR or SZ signals. A technically more
correct way of performing this calculation would be to correctly take
into account the cross-correlation field $\eta$.

We use the linear bias function (we take the simple linear bias of
\cite{1996MNRAS.282..347M}) and $M_{\rm \min}$ is the minimal mass
corresponding to halos of virial temperature of $2 \times 10^4$K.

\subsection{Star-formation history and reionization}

To complete our calculation, we need to also specify the
star-formation rate and reionization history.  In early universe,
massive enough halos accrete and cool gas to form stars. A halo of
mass $M$ converts a fraction of $\fstar\fb$ of its mass into first
massive stars that later explode as supernovae, where $\fstar\sim0.2$
is the fraction of baryons that fragment to form stars and
$\fb=\Omega_{\rm b}/\Omega_{\rm m}\approx 0.2$ is the baryon fraction.

The total density in stars at any redshift is given by
\begin{equation}
\label{eq:51}
\rho_{\star}(z)=  
\int_{\Mmin}^{\infty} \rho_{\star}(M,z)\rmd M =
\int_{\Mmin}^{\infty} \fb \fstar M \frac{\rmd
  N}{\rmd  M} \rmd M ,  
\end{equation}
where $\rmd N/\rmd M$ is the number density of halos per unit mass
range. We take the Press-Schechter \citep{1974ApJ...187..425P} form of
the number counts. The star-formation rate per comoving volume for
halos with mass between $M$ and $M+\rmd M$ is then simply

\begin{equation}
  \label{eq:sfr}
%  \Psi(z) = \frac{\rmd z}{\rmd t} \frac{\rmd}{\rmd z} \rho_{\star}
  \Psi(M,z) = \frac{\rmd}{\rmd t} \rho_{\star} (M,z) = \frac{\rmd
    z}{\rmd t} \frac{\rmd}{\rmd z} \rho_{\star} (M,z)
\end{equation}
with 
\begin{equation}
  \label{eq:dzdt}
  \frac{\rmd z}{\rmd t} = H_0 (1+z) \left(\Omega_{\rm m}(1+z)^3+\Omega_{\rm k}(1+z)^2+ \Omega_\Lambda \right)^{1/2}
\end{equation}

\begin{figure}
  \centering
  \includegraphics[width=1.0\linewidth]{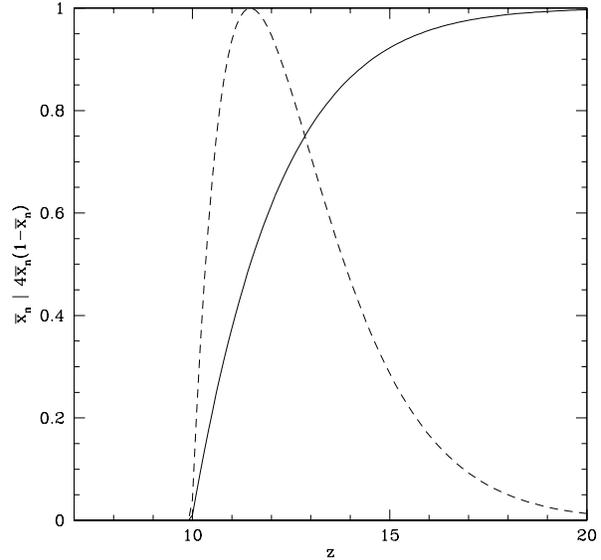}
  \caption{The neutral fraction ($\bhx$) as a function of redshift as
    estimated by the equation (\ref{eq:neutralfraction}). The dashed
    line shows the  $4\bhx(1-\bhx)$ -- a quantity that is directly
    proportional to amplitude of cross-correlation power spectra.
}

 \label{fig:xin}
\end{figure}

We can also make a very simple model that connects the reionized
fraction and star-formation activity. If we take a steady-state
approximation and assume that $q\sim4000$ ionizing photons are emitted per
baryon in stars, then the ionised fraction is simply given by
\begin{equation}
  1-\bhx = {\rm min} \left[ \frac{q t_{\rm rec}}{\Omega_{\rm
        b}\rho_{\rm crit}} \frac{\rmd \rho_\star}{\rmd t} , 1 \right],
  \label{eq:neutralfraction}
\end{equation}
where the recombination time scale is given by \cite{Madau:1998cd}
\begin{equation}
  t_{\rm rec} =  0.3 \left(\frac{\omega_b}{0.02}\right)^{-1}\left(\frac{1+z}{4}\right)^{-3}
  \left(\frac{C}{30}\right)^{-1}
\end{equation}
where we assumed the clumping factor $C=\langle n_{\rm
  HII}^2\rangle/\langle n_{\rm HII}\rangle^2 \sim 50$.
The resulting neutral fraction is plotted in the Figure \ref{fig:xin}.

\subsection{Approximations}

Two common approximations are used in the calculation of power spectra,
depending on the ratio of survey depth compared to the angular scales
of interest. If $\delta r/(D_A \ell^{-1})\ll 1$, then window function
can be approximated as a delta function and the equation \eqref{eq:nn}
reduces to 
\begin{equation}
  C_\ell^{nn} = 4 \pi T_0(r_0)^2 \int \frac{\rmd k}{k} \Delta_{\psi\psi}^2(k)
  j_\ell^2(kr_0),
\label{eq:nnd}
\end{equation}
and equivalent for equations ~\eqref{eq:ii} and \eqref{eq:x}.

The other limit is when $\delta r/(D_A \ell^{-1})\gg1 1$ one can use
the so-called Limber approximation:
\begin{equation}
  \int \rmd k F(k) j_\ell(kr_1) j_\ell (kr_2) \approx
  \frac{\pi}{2r_1^2} \delta(r_1-r_2) F(\ell/r_1)
\end{equation}
The necessary condition is that $F(k)$ is a smoothly varying
function. When applied to the power spectra, this results in 
equation \eqref{eq:nn} simplifying to
\begin{equation}
    C_\ell^{nn} = \int \rmd r (T_0(r) W(r))^2 P(\ell/d_A) r^{-2}.
\label{eq:nnl0}
\end{equation}
If additionally $\delta r\ll r$, one can simply further to 
\begin{equation}
    C_\ell^{nn} = T_0^2(r) P(\ell/d_A) r^{-2} (\delta r)^{-1},
\label{eq:nnl}
\end{equation}
where $\delta r$ is the effective width of the windows function
$W(r)$.
Analogous expressions hold for other equations for cross-correlation
spectra  are:

\begin{eqnarray}
    C_\ell^{ni} = T_0(r) \frac{\rmd I_\nu}{\rmd r (r)} P(\ell/d_A) r^{-2}, \\
    C_\ell^{ny} = T_0(r) \frac{\rmd y}{\rmd r (r)} P(\ell/d_A) r^{-2}.
\label{eq:nnlx}
\end{eqnarray}

The interesting point here is that the auto-correlation power spectrum
in units of temperature scales inversely with bandwidth ($\delta r$), while the
cross-correlation power spectra are, in the Limber limit, to first
order independent of it.

%%%%%%%%%%%%%%%%%%%%%%% FIGURES

\begin
  {figure}
  \centering
  \includegraphics[width=\linewidth]{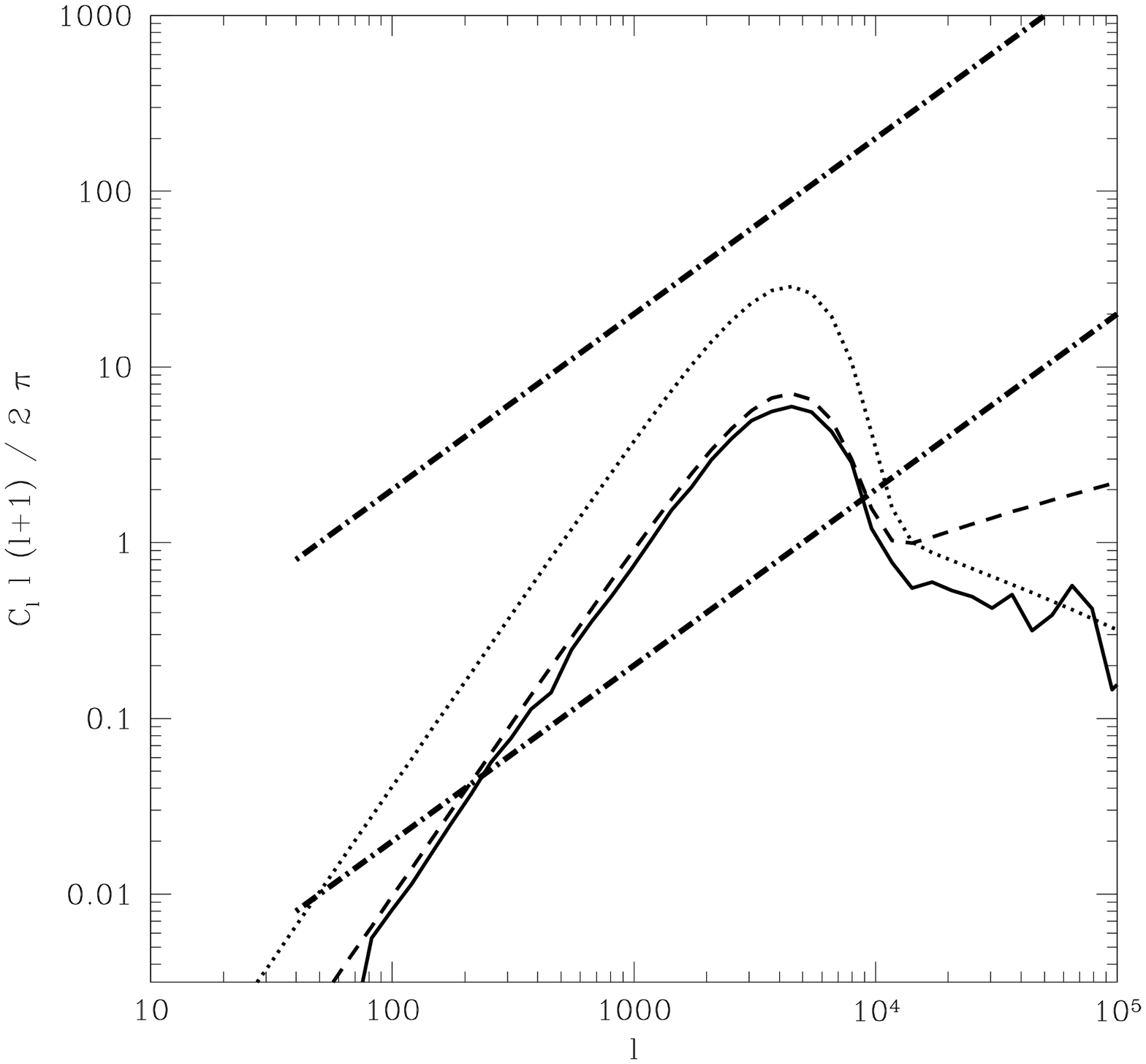}
  \includegraphics[width=\linewidth]{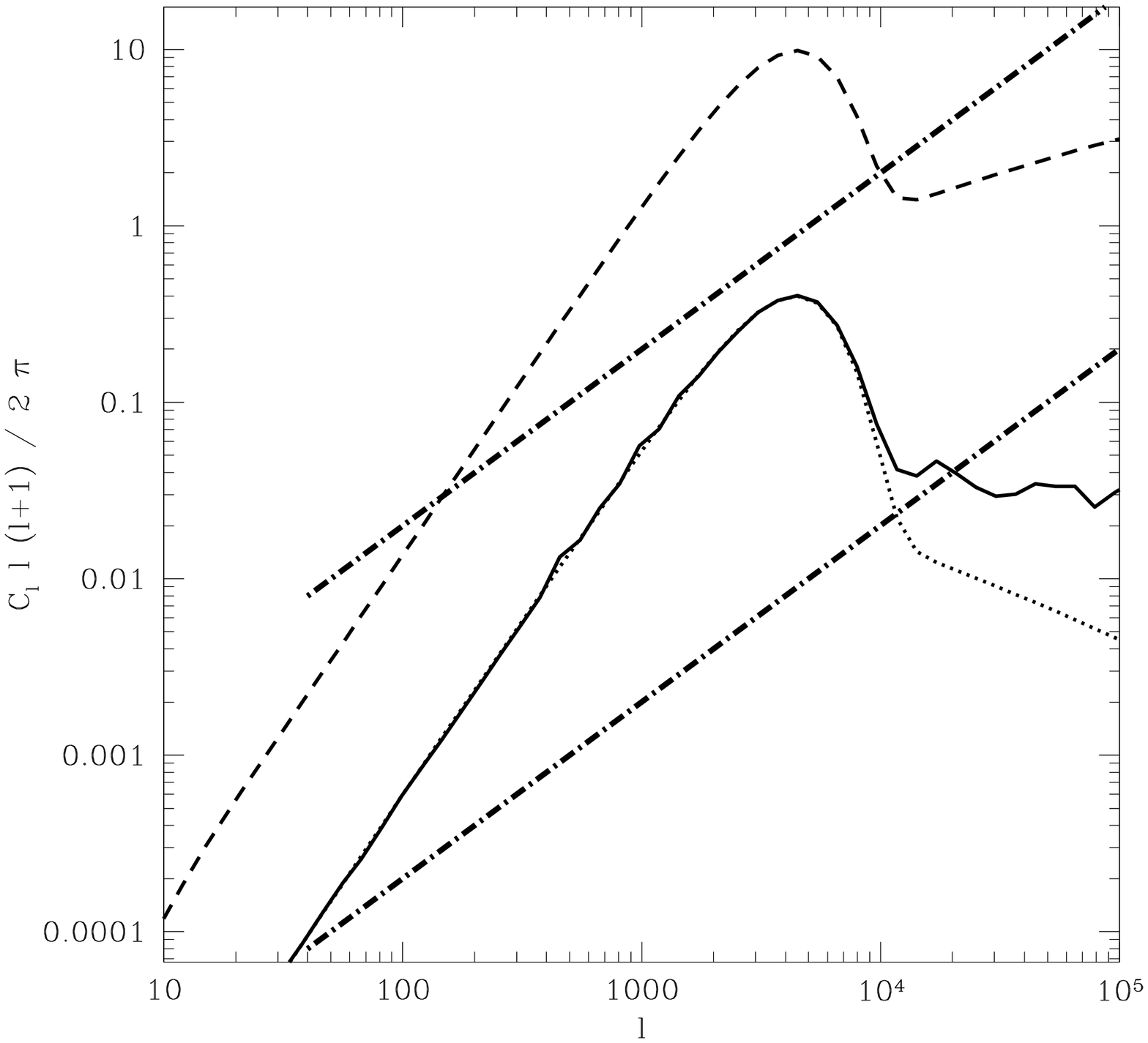}
  \caption{The power spectra calculated for the reionization and bubble
    model used in this paper at $z=10$ using three different
    approximation schemes discussed in the text. Solid line is the
    full calculation of the equation \eqref{eq:nn}, dashed is the
    delta-function approximation of equation \eqref{eq:nnd} and dotted
    line is the Limber approximation of equation \eqref{eq:nnl}. The
    bandwidth of 0.1MHz was used for the top graph and 10MHz for the
    bottom one. The dot-dashed lines denote the expected $\Delta^{\rm noise} C_\ell$ for
    LOFAR and SKA experiments.
}
 \label{fig:which2}
\end{figure}

\begin
  {figure}
  \centering
  \includegraphics[width=\linewidth]{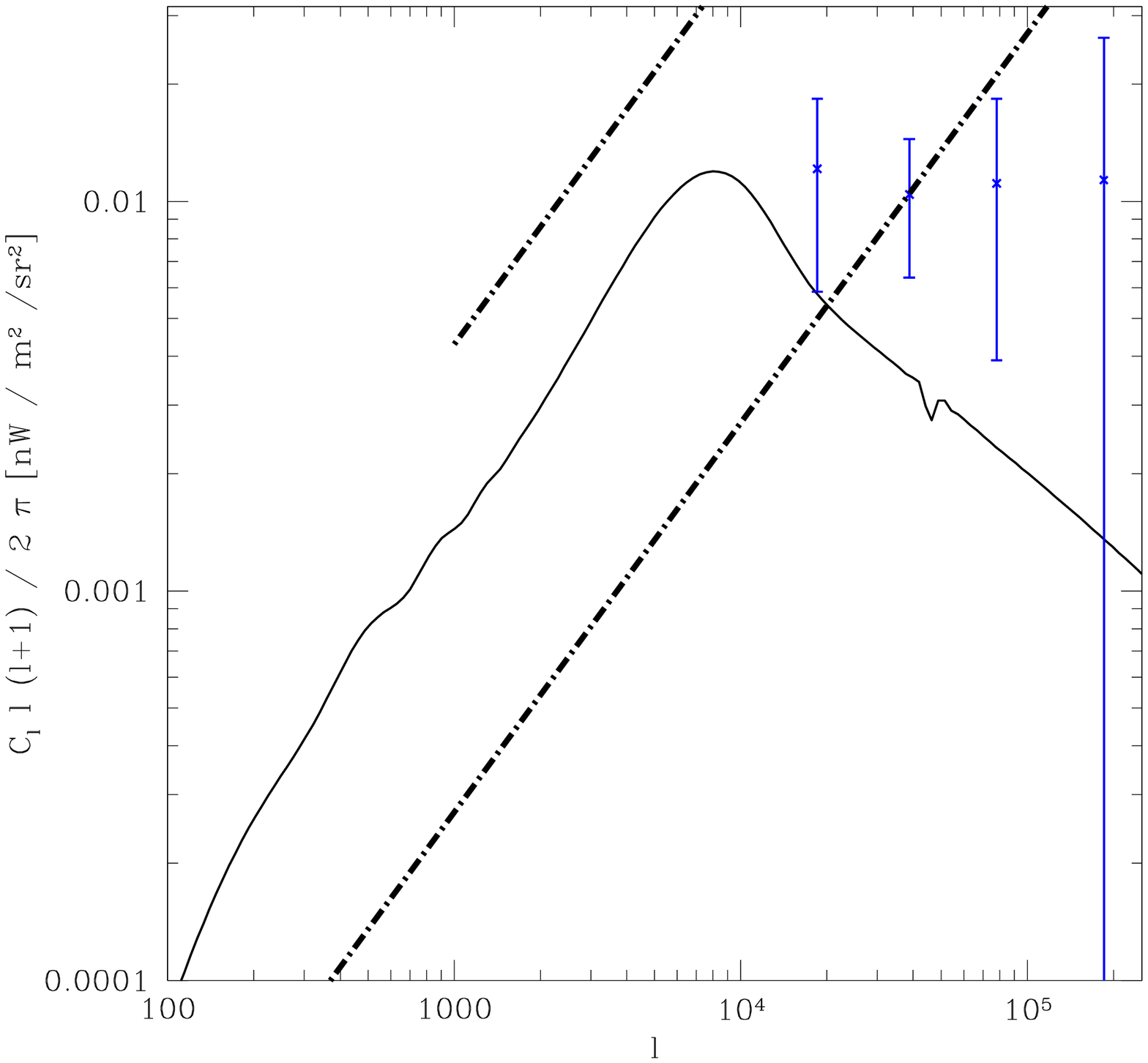}
  \includegraphics[width=\linewidth]{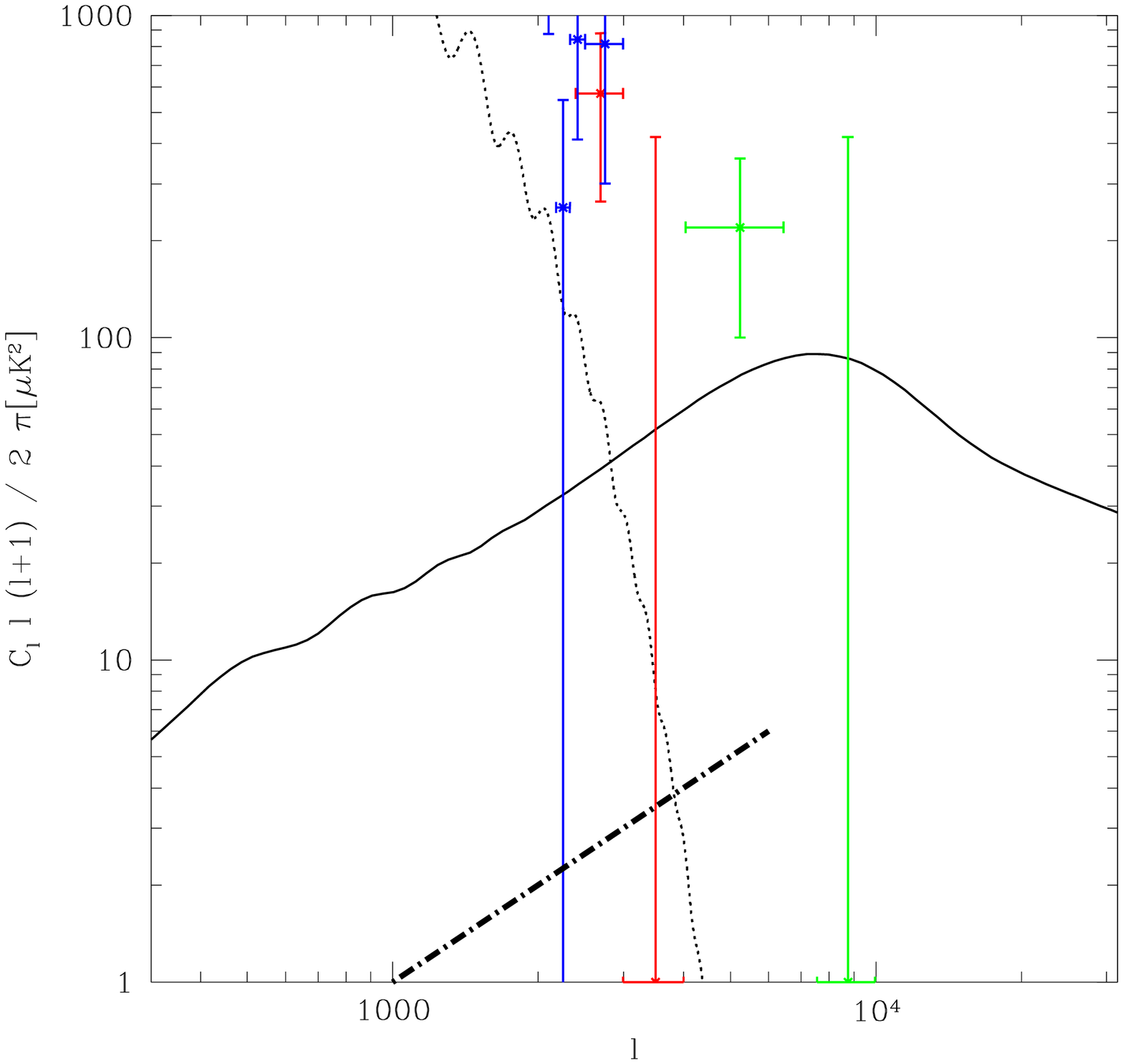}
  \caption{The auto-correlation power spectra contribution from early
    stars for IR (top) and SZ (bottom). For IR contribution we show
    the predicted power spectra, the data-points from Kashlinsky with
    shot-noise contribution subtracted \citep[for details see][]{Sullivan:2006vr} for
    details). Straight dot-dashed lines are the expected $\Delta^{\rm
      noise} C_\ell$ for the CIBER (top; see
    \citep[top; for details see][]{2006NewAR..50..215B} and AKARI
    \citep[for details see][]{Matsuhara:2006bp} experiments. For SZ
    contribution we plot the predicted auto-correlation power
    spectrum, the data-points from CBI (red), ACBAR (blue)
    (\emph{rescaled by 4.3 to account for different observing
      frequency of ACBAR assuming that the signal is dominated by the
      SZ effect}) and BIMA (green). The straight dot-dashed line is
    the expected $\Delta^{\rm noise} C_\ell$ for the PLANCK
    experiment.
    \label{fig:auto} }

\end{figure}

%%%%%%%%%%%%%%%%%%%%%%%%%%%%% END FIGURES 

%%%%%%%%%%%%%%%%%%%%%%%%%%%%%%%%%%%%%%%%
\begin{figure*}
  \centering
  \begin{tabular}{cc}
  \includegraphics[width=0.5\linewidth]{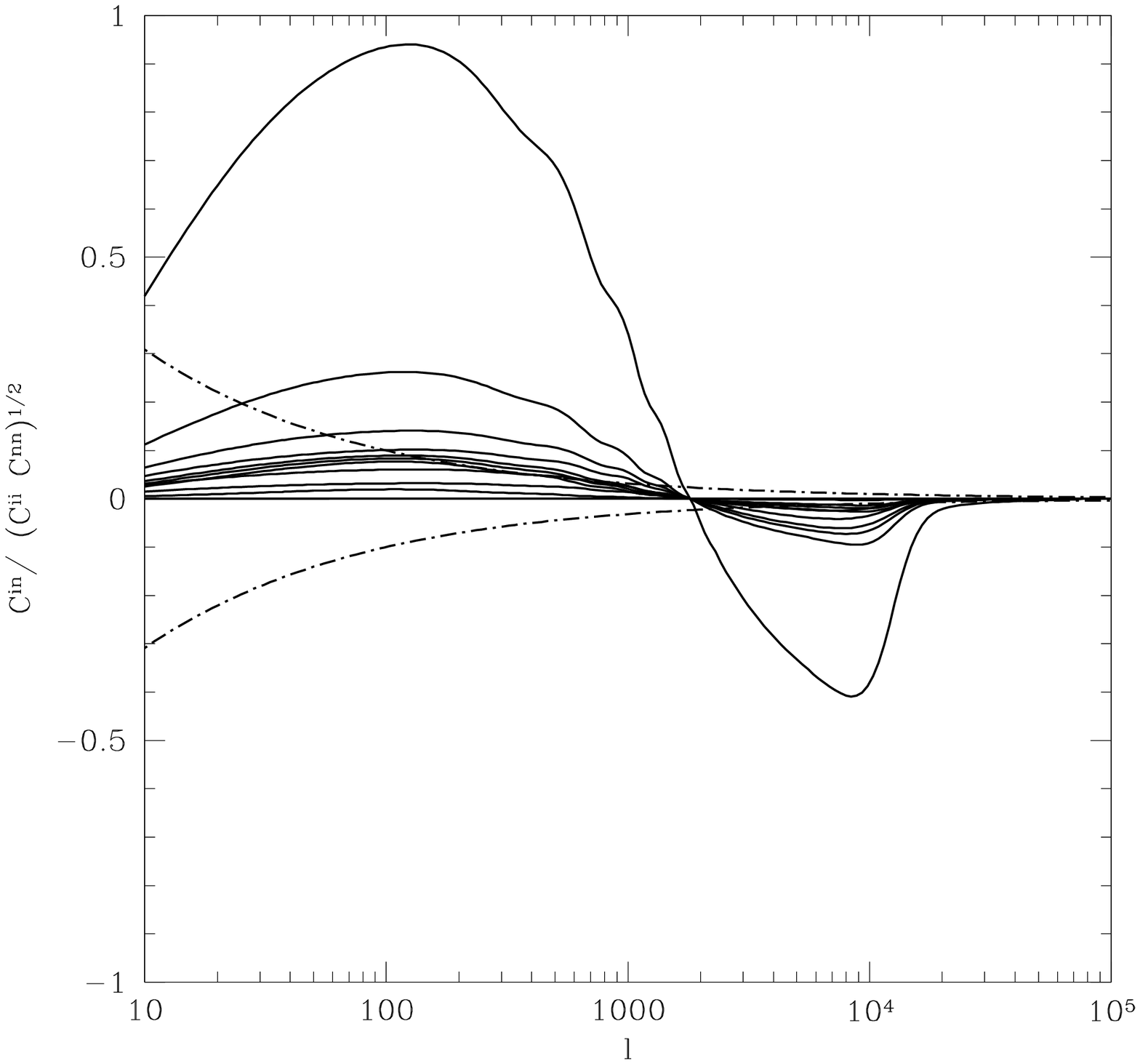} &
  \includegraphics[width=0.5\linewidth]{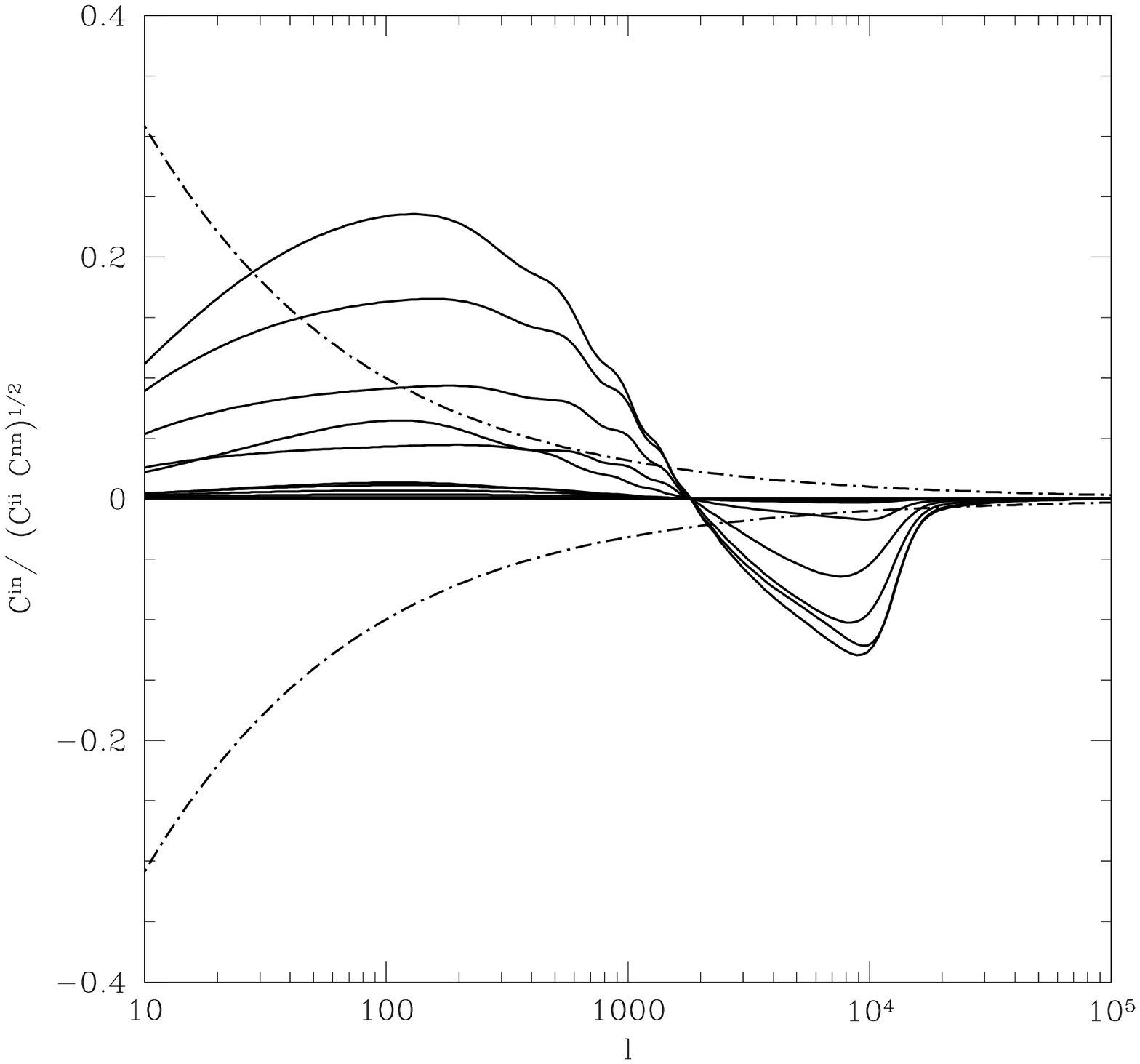} \\
  \end{tabular}

  \caption{The cross-correlation power spectra for 21-cm anisotropy at
    redshifts between 8 and 19 correlated against IRB observed at
    wavelengths of $1.94\mu$m (left panel; frequency chosen so that
    peak occurs at integer redshift, see also Figure
    \ref{fig:source}) and $4\mu$m (right panel).  The bandwidth of
    the 21-cm signal was set to 10MHz.  The redshift of maximum
    amplitude is 15 for $1.94\mu$m and 11 for for $4\mu$m. }
 \label{fig:crossrel}
\end{figure*}

%%%%%%%%%%%%%%%%%%%%%%%%%%%%%%%%%%%%%%%%%%%%%%%%%

% bandwidth.  
In Figure \ref{fig:which2}, we plot a sample spectrum using the full
formula \eqref{eq:nn} and the two approximations in equations
\eqref{eq:nnd} and \eqref{eq:nnl}. We obtain the expected result: for
narrow bandwidths and large angular scales the delta function
approximation fares better, but for bigger bandwidths the Limber
approximation is more appropriate. The unexpected discrepancy at large
bandwidths and small scales is likely due to ringing of the window
function, which does not seem to cancel out completely as assumed in
the Limber approximation.

\section{Results}
\label{sec:results}

The auto-correlation power spectra of our model are comparable with
published results. We plot some of them, together with some published
data in the Figure \ref{fig:auto}. The observational uncertainties in
the parameters of our model give an overall uncertainty in the
amplitude of the signal of about order of magnitude. The mean
Comptonization parameter in our model is $\bar{y}=5\times 10^{-6}$,
which is compatible with the COBE-FIRAS upper limit around at the same
level \citep{1994ApJ...420..439M}. The 2$\mu$m mean IR flux from first sources
during reionization $\sim$ $5$ nW m$^{-2}$ sr$^{-1}$ an 2$\mu$m, which
is within experimental upper limits on the IRB intensity (about $20$
to 70 nW m$^{-2}$ sr$^{-1}$) at those wavelengths \citep{Cambresy:2001ei}.

\begin{figure}
  \centering
  \includegraphics[width=\linewidth]{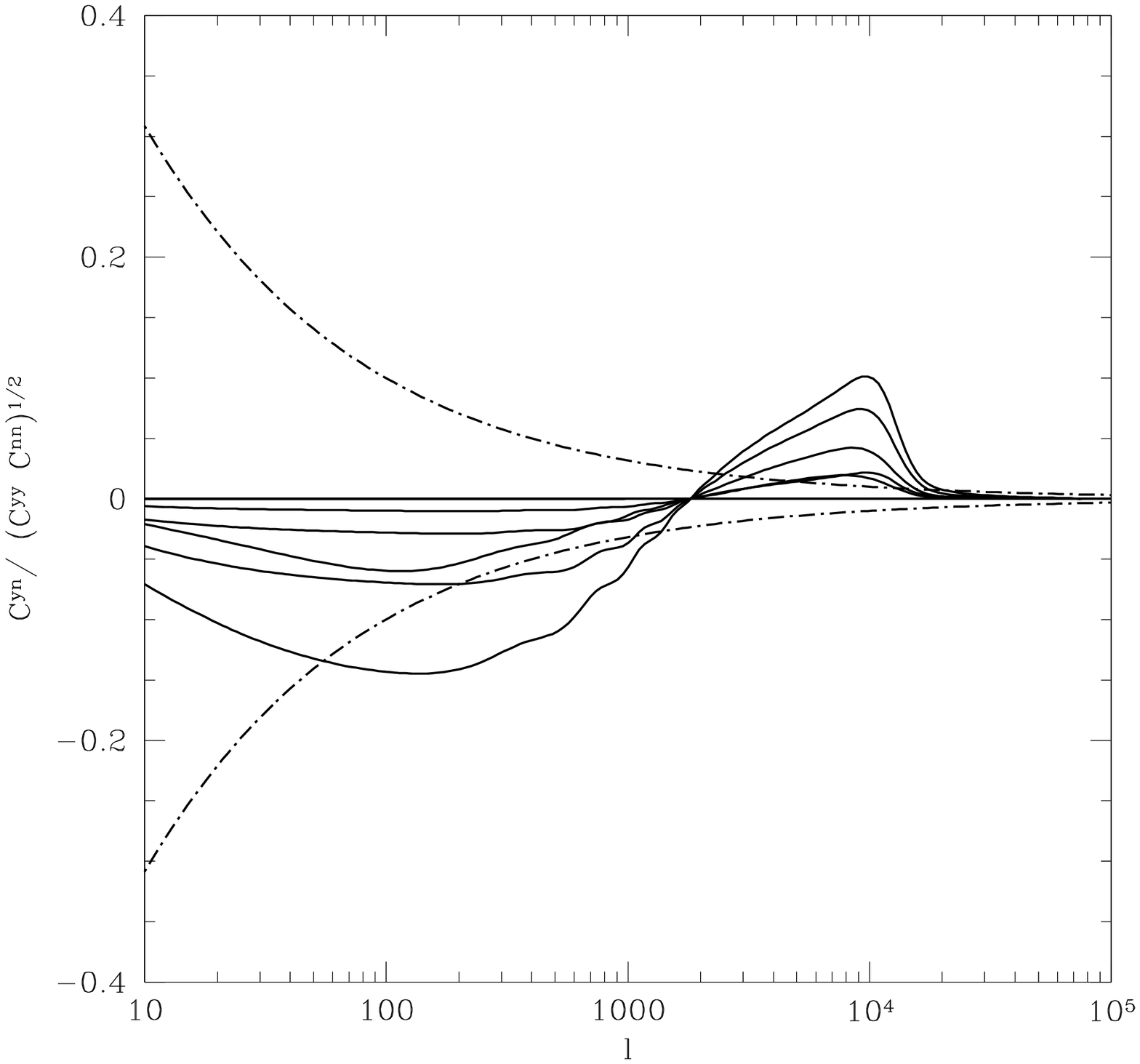} 
  \caption{The same as figure \ref{fig:crossrel} but for
    cross-correlating 21-cm with Comptonization parameter expressed in
  units of temperature decrement in the Rayleigh-Jeans region of the spectrum.}
 \label{fig:crossrel2}
\end{figure}

In Figure \ref{fig:crossrel} we plot the relative cross-correlation
coefficient $C_\ell^{ni}/\sqrt{C_\ell^{ii} C_\ell^{nn}}$ for the
cross-correlation of the 21cm and the IR signals.  We naturally expect
a negative cross-correlation between the red-shifted 21-cm signal and
the infrared light from the stars. Even in our simplified model this
anti-correlation does not hold on large scales where the signal is
dominated by the density perturbations that correlate positively. The
cross-over scale correspond to the typical size of the bubbles.
Therefore, we generically predict that IR and 21-cm line signals to be
correlated on large scale and anti-correlated on small scales. The
cross-over scale can be used to determine the typical size of the
bubbles. Our calculations are probably accurate to only an order of
magnitude, but the heuristics should nevertheless be robust.
Secondly, we note that when small enough ($\sim 2\mu$m) wavelength is
used, then the IR signal is dominated by a rather narrow redshift
range and the cross-correlation will be picked only when 21-cm
observation are tuned to the same redshift range. When observing the
continuum radiation at larger wavelengths, the cross-correlation is
much less pronounced, but present at all redshifts. In Figure
\ref{fig:crossrel2} we plot the same for cross-correlating the 21-cm
signal with the SZ signal. We note that this graphs looks suspiciously
similar to the right panel of Figure \ref{fig:crossrel}.

\begin{figure}
  \centering
  \includegraphics[width=\linewidth]{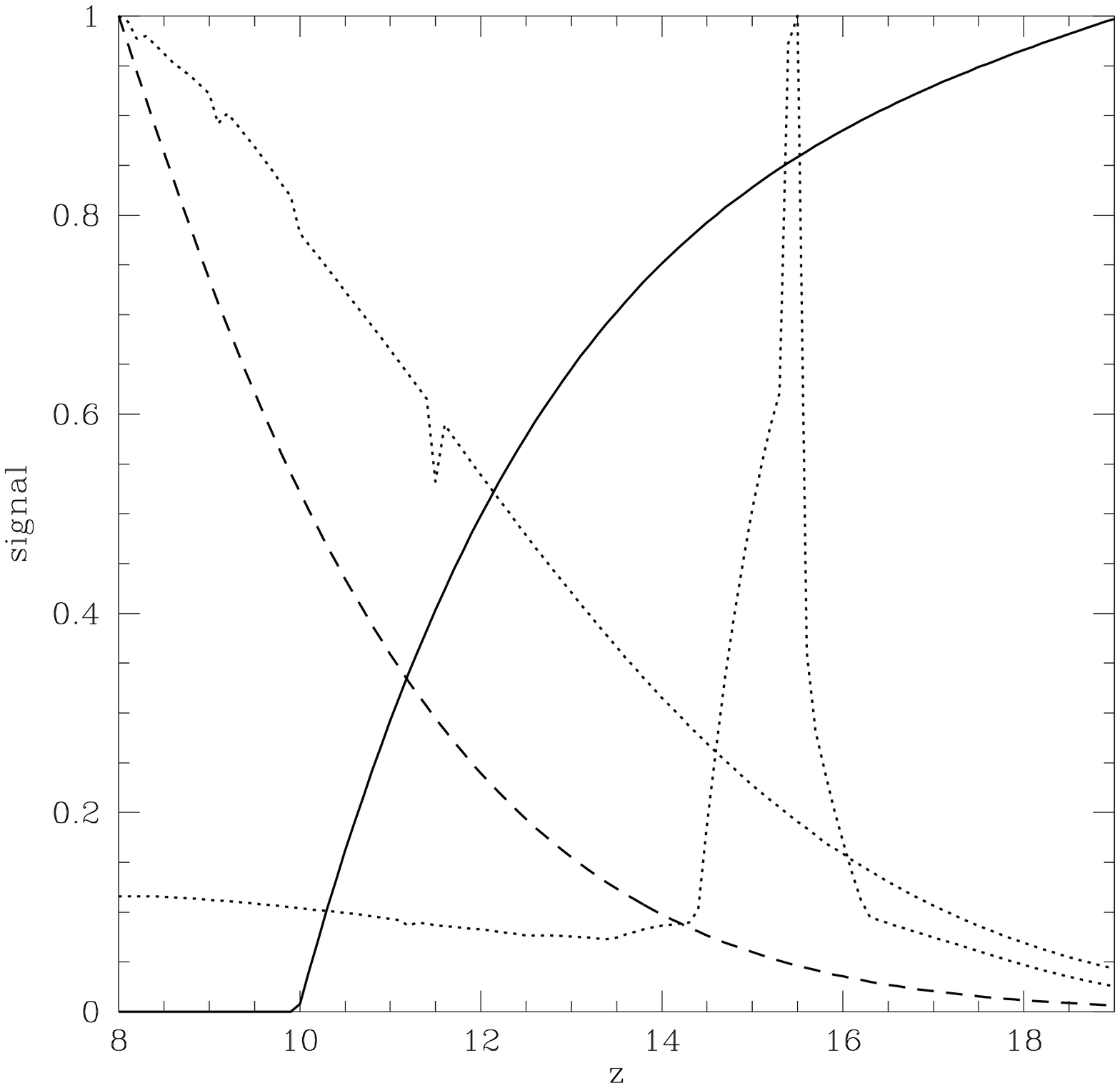} 

  \caption{ This figure shows the functions that source auto- and
    cross- correlation power spectra as a function of redshift. We
    plot $T_0 \bar{\psi}$ (solid line), $\rmd I_\nu/\rmd r \bar{\phi}$
    (for $\nu_{\rm obs}=2\mu$m which has spike at $z\sim 15$
    and $\nu_{\rm obs}=4\mu$m; dotted) and $\rmd y/\rmd r \bar{\phi}$. All
    curves were normalized to unity at maximum.}

 \label{fig:source}
\end{figure}

These features are easy to understand, if we look at the ``source''
functions plotted in Figure \ref{fig:source}. There we plot the
functions which are proportional to the amplitude of the signal given
the same power spectrum. We note that particularly for the case of the
IR signal, we can clearly see the features of the spectrum plotted in
the Figure \ref{fig:pop3}: If we choose to observe the spectrum at
large wavelengths, the spectrum is essentially featureless, while if
we choose to observe at $\sim 2\mu$m we get the pronounced spike at
$z\sim 15$. It is clear from this graph, the the maximum
cross-correlation for either SZ or IR signal at 4$\mu$m should happen
at $z\sim 11$, while the IR correlation is nearly completely sourced at
$z \sim 15$.

\subsection{Observability}

Is the signal large enough to be realistically observed by any of the
future experiments?

There are two sources of noise present in any attempt to detect
cross-correlation between the two fields: the intrinsic noise
properties of instruments used and the sample variance that
limits our ability to detect correlations due to a finite solid angle
of the sky being observed.

The noise power spectrum for the planned future radio telescopes can
be, under some simplifying assumptions be written as
\citep{Zaldarriaga:2003du}
\begin{equation}
\frac{\ell^2 C_\ell^{\rm noise}}{2\pi} = \frac{T_{\rm sys}^2 (2 \pi)^2}{\delta
  \nu t_{\rm int} f^2_{\rm cover}} \left(\frac{\ell}{\ell_{\rm max}}\right) 
\end{equation}
where $T_{\rm sys}$ is the system temperature, $\delta \nu$ is the
experiment bandwidth, $\delta \nu$ is the experiment bandwidth,
$t_{\rm int}$ is the integration time and $f_{\rm cover}$ is the
covering fraction.  The error on the estimated power spectrum is then
given by 
\begin{equation}
 \Delta^{\rm noise} C_\ell^{nn}   \sim C_\ell^{\rm noise} \frac{\ell_{\rm min}}{\ell}
\end{equation}

Very similar consideration hold for IR and SZ experiments. The noise
power spectrum is given by
\begin{equation}
  C_\ell^{\rm noise} = 4\pi f_{\rm sky} \frac{\sigma^2}{N_{\rm pix}},
\end{equation}
where $f_{\rm sky}$ is the fraction of sky observed (or being observed
by both experiments for cross-correlation studies), $N_{\rm pix}$ is
the number of pixel and $\sigma$ is noise per pixel. The error on the
estimated power spectrum is then given by
\begin{equation}
\Delta^{\rm noise} C_\ell= \sqrt{\frac{2}{f_{\rm sky} (2\ell+1)}} C_{\ell}^{\rm noise}
\end{equation}

In the Table \ref{tab} we list the expected sensitivities for a few
experiments. We discuss the following experiments in this paper:
Low Frequency Array \citep[LOFAR;][]{2004AAS...20515308S}\footnote{\texttt{http://www.lofar.org}},
the Square Kilometer Array \citep[SKA;][]{2004astro.ph..9274C}\footnote{\texttt{http://www.skatelescope.org}},
CIBER \citep{2006NewAR..50..215B}\footnote{http://physics.ucsd.edu/~bkeating/CIBER.html},
AKARI \citep{Matsuhara:2006bp}\footnote{\texttt{http://www.ir.isas.jaxa.jp/ASTRO-F/Outreach/index\_e.html}}
and 
PLANCK \citep{2006astro.ph..4069T}\footnote{\texttt{http://www.rssd.esa.int/index.php?project=Planck}}.

\begin{table*}

  \begin{tabular}{ccccc}
    Experiment & Type & $\ell_{\rm min} - \ell_{\rm max}$ & $f_{\rm sky}$ &  $\Delta^{\rm noise}  C_\ell \ell^2/2\pi$ at $\ell=10^3$ \\ %& $\Delta^{\rm SV}  C_\ell \ell^2/2\pi$ at $\ell=10^3$  \\
\hline
    LOFAR & 21-cm & $\sim$ 40 - $\sim 10^4$ & $\sim 0.5$    &   $0.2$mK$^2$ \\%& $0.08$mK$^2$ \\
    SKA & 21-cm  &  $\sim$ 40 - $\sim 10^4$  & $\sim 0.5$ & $0.002$mK$^2$  \\%& $0.08$mK$^2$  \\
\hline
    CIBER & IR & $\sim 10^3$ - $\sim 10^4$  & $4 \times 10^{-4}$ & $0.004$  (nW/m$^2$/sr) \\%& $6$  (nW/m$^2$/sr)   \\
    AKARI & IR & $\sim$ 2 - $5\sim 10^5$  & 1  & $3 \times 10^{-4}$  (nW/m$^2$/sr) \\%& $0.1$ (nW/m$^2$/sr  \\
\hline
    Planck  & SZ & $\sim$ 2 - $\sim 6000$  & $\sim    1$ & $1 \mu {\rm K}^2$ \\ %%& 0.1 \mu {\rm K}^2$   \\ 
  \end{tabular}

  \caption{This table lists the expected sensitivities for two 21-cm
    and one IR and CMB experiments. For 21-cm signal we assumed one
    month observation and 10MHz bandwidth (taken from
    \citep{Zaldarriaga:2003du}). For the rocket IR experiment, we
    assumed 1.8nW/m$^2$/sr per pixel with pixel size 17'' and 16
    square degrees of observation and for the fictional AKARI FS (full
    sky) experiment we assumed noise of 11 $\mu$Jy, beam size of 1.46 arc mins,
    corresponding to the NEP (North Ecliptic Pole) survey, but
    expanding it from 6 square degrees to full
    sky. For Planck experiment we assumed full 
    sky observation at 10 arcmin resolution and noise per pixel of 5
    $\mu {\rm K}^2$.  \label{tab}}
\end{table*}

% Finally, we assume that the error on the cross power spectrum is
% simply given by
% \begin{equation}
%   \delta C_\ell^{ni} = \sqrt{\delta C_\ell^{ii} \delta C_\ell^{nn}}
% \end{equation}

\begin{figure*}
  \centering
  \begin{tabular}{ccc}
  \includegraphics[width=0.33\linewidth]{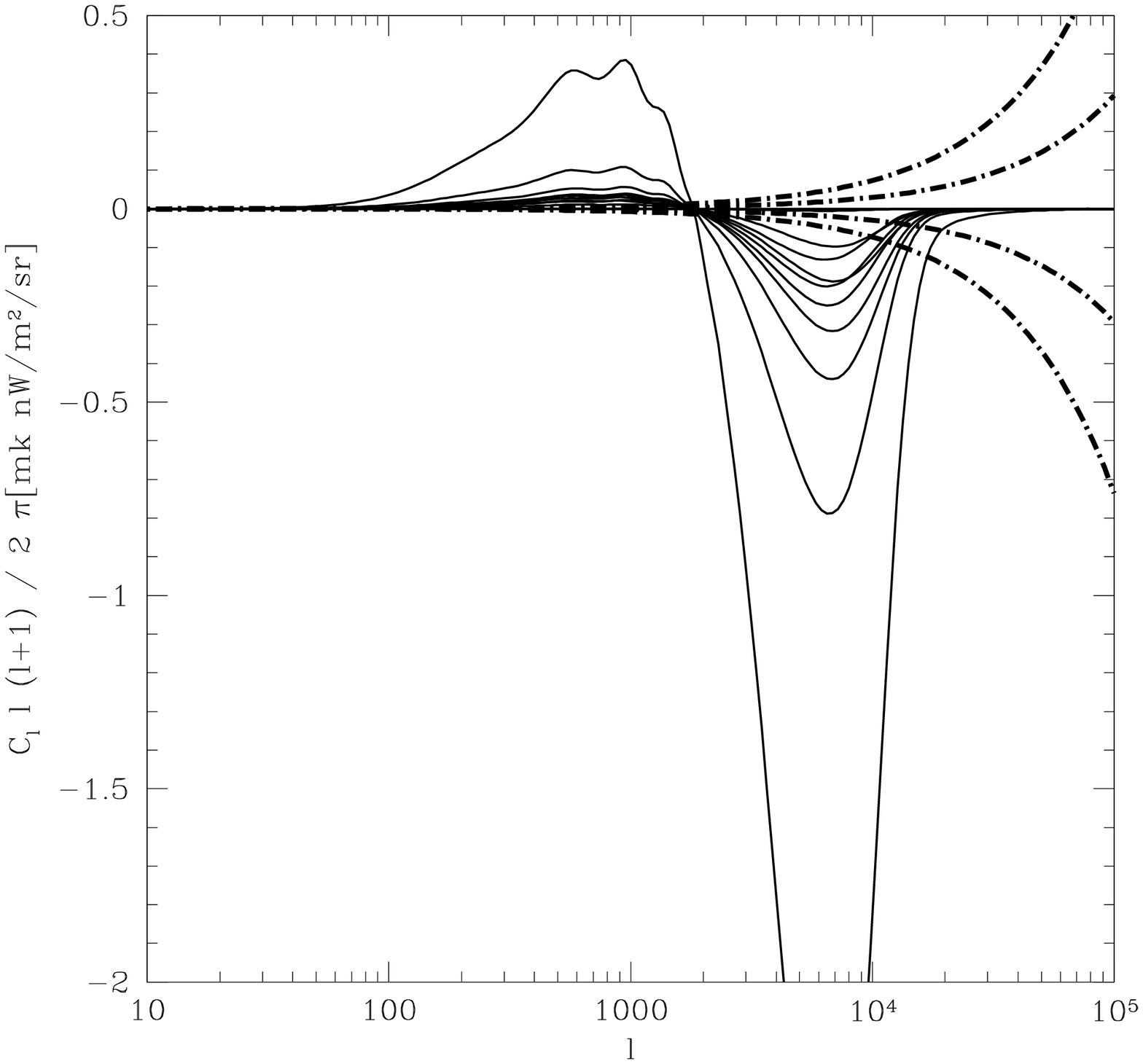} &
  \includegraphics[width=0.33\linewidth]{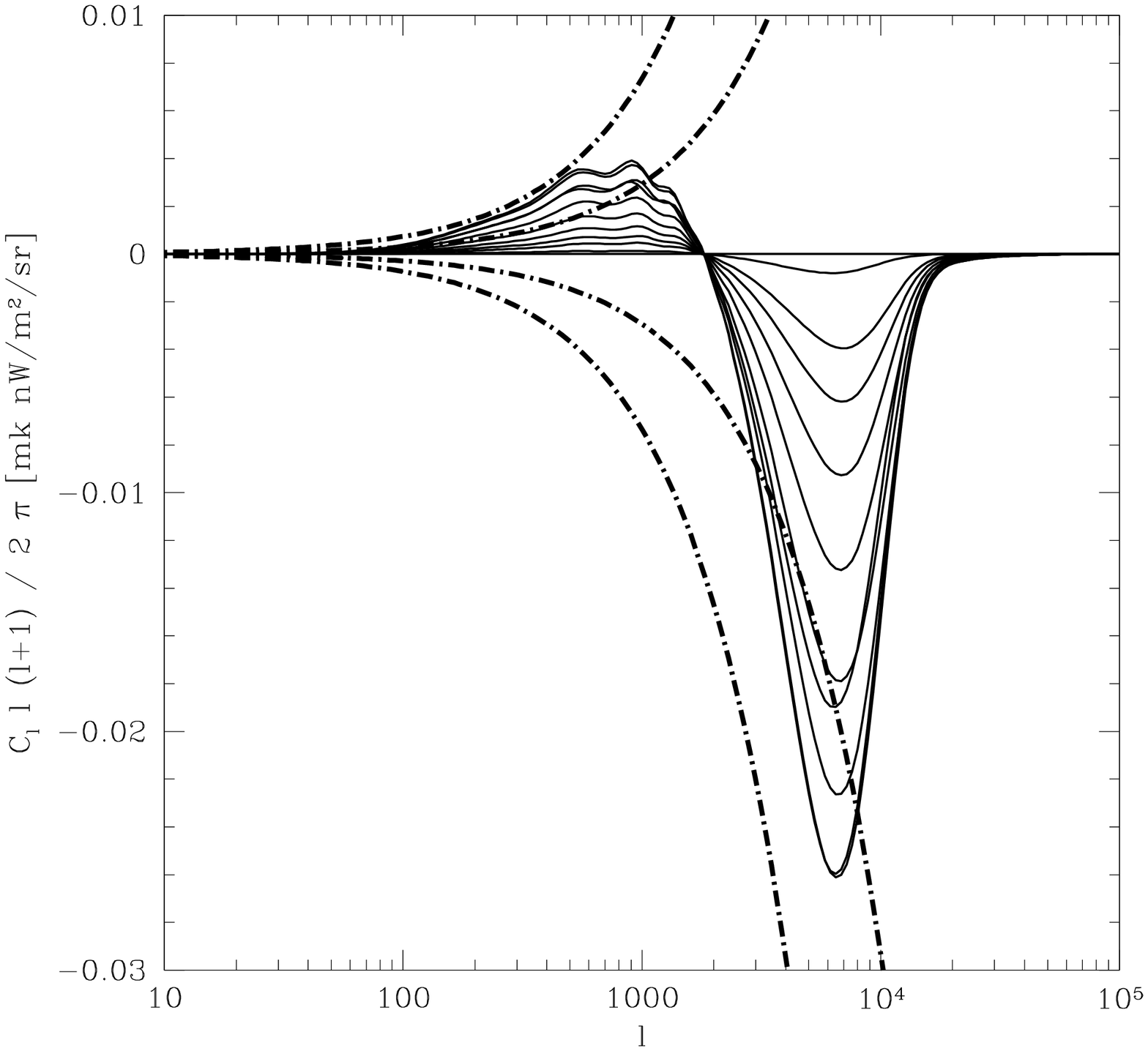} &
  \includegraphics[width=0.33\linewidth]{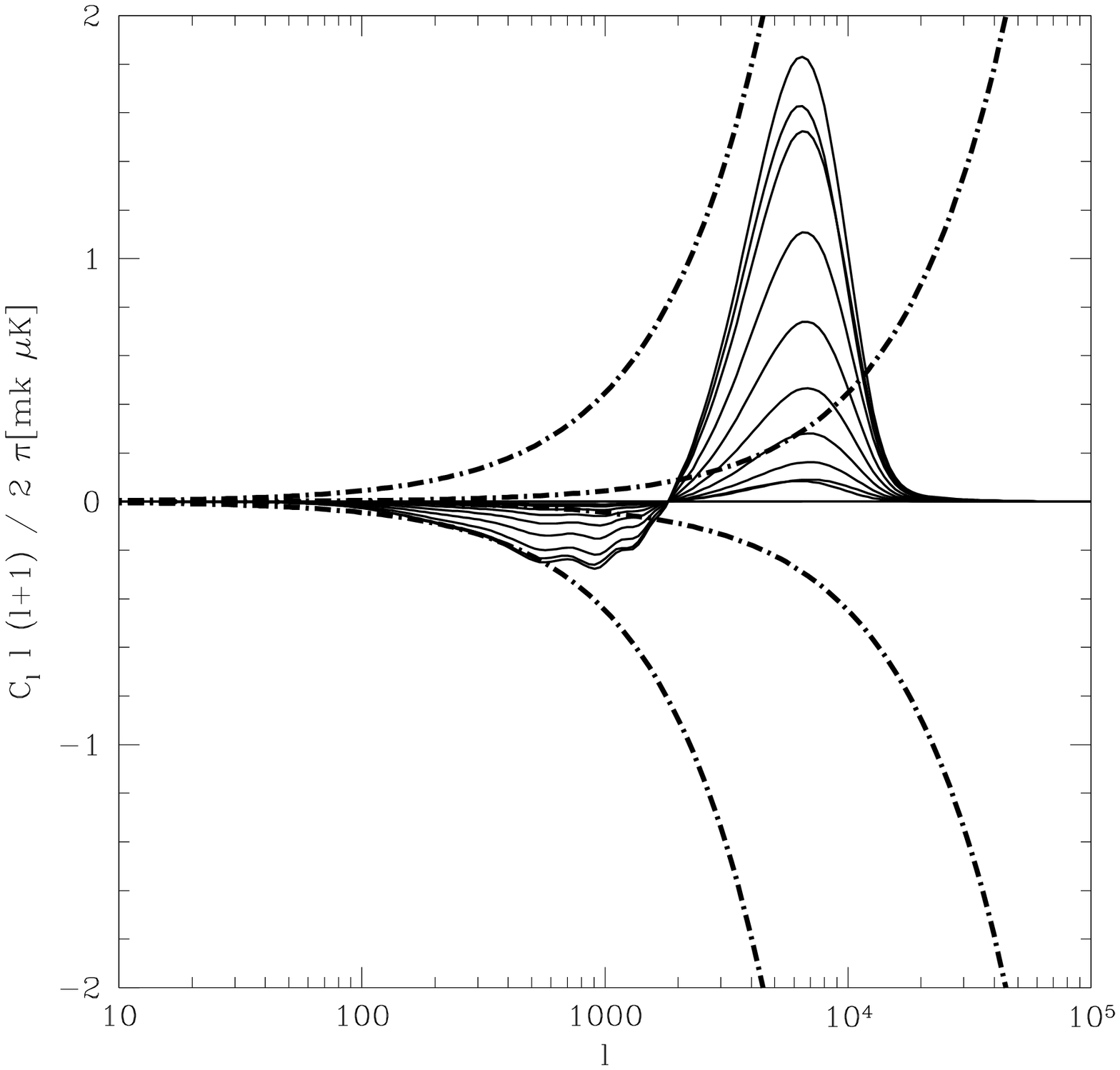} \\
  \end{tabular}

  \caption{This figure shows the same cross-correlations as in the
    Figures \ref{fig:crossrel} and \ref{fig:crossrel2}, but in
    absolute units. Left and middle panels are for 21-cm -IR cross
    correlation at $1.94\mu$m and 4$\mu$m respectively, the right panel
    is for 21-cm-SZ cross-correlation.  The expected sensitivities for
    experiments in Table \ref{tab} are also plotted as dashed lines
    (outer contours are for cross-correlations with LOFAR, inner for
    cross-correlations with SKA. \label{fig:cross} }
\end{figure*}

In Figure \ref{fig:cross}, we plot predictions for cross-correlation power spectra and the
instrumental sensitivities.

However, sample variance is not necessarily a negligible problem in
cross-correlation studies or cases with small $f_{\rm sky}$: only a
small part of the IR or SZ light cross-correlates with the relatively
narrow-band 21-cm light and so the rest of the IR signal acts as a
source of noise. As a crude approximation, we can assume that the
skies are normally distributed and so the sample variance errors are
given by
\begin{equation}
  \Delta^{SV} C_\ell^{xy} = \sqrt{\frac{2 C_\ell^{xx}
      C_\ell^{yy}}{f_{\rm sky} (2 \ell+1})},
\end{equation}
for $\ell \gg 2$ and the dummy indices $x$ and $y$ are to be replaced
with $n$ (21-cm signal), $i$ (IR signal) or $y$ (SZ signal).
%  Quantity
% $f_{\rm sky}$ is the fraction of sky being observed (or observed by
% both experiments for the case of cross-correlation). 
For $f_{\rm
  sky}=1$ the sample variance is often termed cosmic variance.

This justifies our choice of bandwidth to be $10$ MHz. When entering
the Limber regime, the 21cm signal drops in amplitude, while the
cross-correlation signal stays, to first order, the same. Hence, the
cross-correlation is easier to detect if we use wide-bandwidth
experiment (or bin the data in a suitable manner). In Figures
\ref{fig:crossrel} and \ref{fig:crossrel2} we plotted the
cosmic-variance limits as a dashed line. At 10MHz they do no seem to
be a problem, but they might be at smaller bandwidths. One must also
note that these error-bars increase with $f_{\rm sky}^{-1/2}$, where
$f_{\rm sky}$ is the fraction of sky being observed and so this might
be a serious problem for experiments that observe only small patches
of the sky. Note, however, that these lines correspond to errors per
single $\ell$: if one combines a range of $\ell$ values as is usually
the case, the error drops correspondingly.

The total error is just the of sample variance error and the noise
error added in quadrature:

\begin{equation}
\left( \Delta^{\rm tot} C_\ell \right)^2 = \left(\Delta^{\rm SV}
  C_\ell \right)^2 + \left( \Delta^{\rm noise} C_\ell \right)^2 
\end{equation}

\begin{figure}
  \centering
  \includegraphics[width=\linewidth]{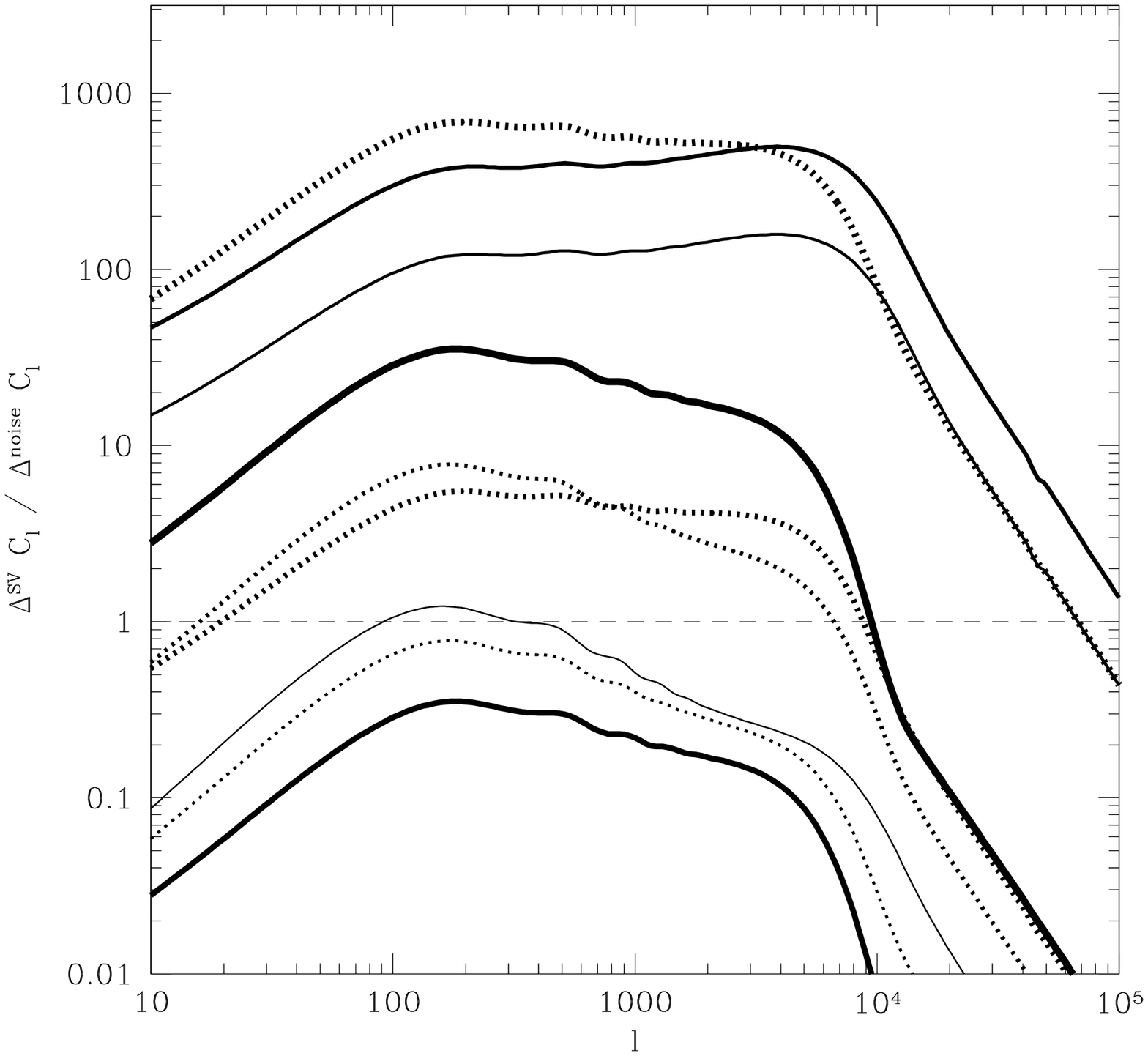} 

  \caption{This figure shows the ratio of sample variance to noise
    error per $C_\ell$. The solid lines are for auto-correlations with
    experiments under consideration are given by line weight: LOFAR
    (thickest), SKA, CIBER, AKARI, Planck (thinnest). The dotted
    line are for cross-correlations and we plot the following combinations: LOFAR $\times$
    CIBER (thickest), SKA $\times$ AKARI, LOFAR $\times$ Planck,
    SKA $\times$ AKARI Planck (thinnest).
    \label{fig:rat} }
\end{figure}

Which source of error is dominant? In the Figure \ref{fig:rat} we plot
the ratio of sample variance to noise error on $C_\ell$ for various auto- and cross- spectra. 
We note that the sample variance is important in most of the
experiments discussed here. This implies that increasing the fraction
of the observed sky will improve constraints.

The overall signal-to-noise ratio (SNR) for a given experiment or pair of
experiments can be simply calculated by

\begin{equation}
  \rm{SNR}^{2} = \sum_\ell \left(\frac{C_\ell}{\Delta^{\rm tot} C_\ell} \right)^{2},
\end{equation}
where index $\ell$ runs over the multipole values for which a given
experiment (or both in case of cross-correlation) has coverage.
Assuming values from Table \ref{tab}, we have estimated the total SNR
for various experiments. These are given in the Table \ref{tab2}.
Note that we chosen $z=14$ as the target 21-cm redshift. If we chose
the redshift corresponding to the peak of the Pop-III spectrum, the
SNR values would have been much more favorable.

\begin{table}

  \begin{tabular}{cc}
    Experiment &  SNR   \\% & SNR excluding SV\\%
\hline
    LOFAR $\times$ LOFAR & $\sim 250$ \\%& $\sim 10^{5}$  \\%
    SKA $\times$ SKA &  $ \sim 2000 $ \\%& $\sim 10^{6}$ \\%
\hline
    CIBER $\times$ CIBER & $\sim 1000$ \\% & $\sim 3000$ \\%
    AKARI FS $\times$ AKARI FS & $\sim 4 \times 10^4$  \\%& $\sim 3000$ \\%
    Planck $\times$ Planck &  $\sim 800 $ \\%& $\sim 800$ \\%
\hline
    LOFAR $\times$ CIBER & $ \sim 40$ \\%& $\sim 38$ \\%
    SKA $\times$ AKARI FS &  $ \sim 1300$ \\%& $\sim 173$\\%
\hline
    LOFAR $\times$ Planck &  $ \sim 36 $ \\%& $\sim 80$ \\%
    SKA $\times$ Planck &  $\sim 137$ \\% & $\sim 350$  \\
  \end{tabular}

  \caption{This table lists the expected signal-to-noise ratios for
    various experiments discussed, assuming the target redshift of
    $z=14$ and observing wavelength $\sim 2\mu$m. 
    \label{tab2}}

\end{table}

\subsection{Recovering $\bhx(z)$}

\begin{figure}
   \centering
   \includegraphics[width=\linewidth]{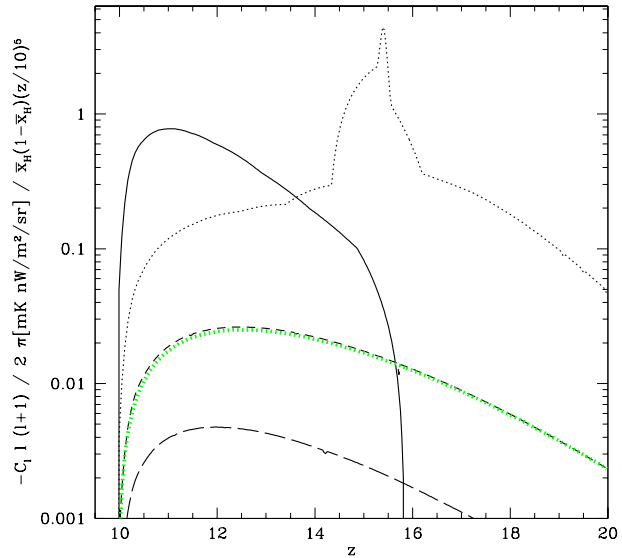} 

   \caption{This figure shows the cross-correlation power spectra 
      at $\ell=6000$ as function of target for IR observing
     wavelengths of 1 $\mu m$ (solid), 2 $\mu m$ (dotted), 4
     $\mu m$ (dashed) and 10 $\mu m$ (long dashed). The thick green dotted
     line in the plot of $\bhx(1-\bhx)(z/11)^2$ }
  \label{fig:abc}
 \end{figure}

From what it was said above, it is clear that there is hope for
reconstruction reionization history using cross-correlations discussed
here.  In Figure \ref{fig:abc} we plot the cross-correlation as a
function of redshift at $\ell=6000$ for a number of observing
wavelengths. The same plot for SZ-21cm correlations would
qualitatively very similar to that of the case of observing at 4$\mu$m
or 10 $\mu$m.  In the same plot we also plot the value of
$\bhx(1-\bhx)(z/11)^2$.

It is clear from the plot that one can, if one observes at the right
wavelength, in principle constrain $\bhx(1-\bhx)$ and therefore the
neutral fraction as a function of time. We will not go into details of
feasibility as there are many theoretical details that needs to be
worked out, in particular the details of star-formation, spectral
features, bubble size, etc.: difficult problems, which should be
nonetheless solved by more detailed modeling. We also note, that
there are several spectra: many degeneracies can be broken by, for
example, using auto-correlation spectra to recover star-formation
histories, etc.

\section{Discussion and Conclusions}
\label{sec:disc-concl}

In this paper we have presented a simple model for auto- and
cross-correlations of the 21-cm, IR and SZ signals from the first
stars.  In particular, the cross-correlations of 21-cm signal with
either IR or SZ signals as a tool to measure reionization history was
introduced for the first time.

The cross-correlation power spectra have a distinct shape: on scales
smaller than the bubble size the anti-correlate, because stars and
supernovae glow in negative with respect to the 21-cm sky. On large scales,
they both trace density fluctuations and correlate positively. 

Cross-correlations have the usual advantage over auto-correlations in
that foregrounds automatically cancel, unless foregrounds for two
signals also cross-correlate, which is unlikely due to a very
different nature of foregrounds for the signals discussed in this
paper. A realistic possibility are radio-galaxy/AGN that can bee seen
both in radio through synchrotron emission and in the infrared by the
dust emission. \cite{DiMatteo:2004dt} have argued that if bright
sources are removed above flux levels of $S\gtrsim0.1$mJy the
\emph{auto}-correlation power spectra should be uncontaminated for
$\ell\lesssim 10^4$ and only contaminated at $O(1)$ for
$\ell\gtrsim10^4$. One could argue that something similar holds for
the cross-correlation. If significant, one could model this
cross-correlation and add it to the model: note that the sign of
cross-correlation is different on small scales and hence this cannot
produce a false detection. Finally, one could resort to using spectral
cleaning techniques \citep{Morales:2005qk}, although this would defy
one of the main advantages of cross-correlation.

We have shown that the tomography of cross-correlations of 21-cm
signal with the IR and SZ signal has, in principle, potential to
strongly constrain the reionization, especially in the region when
$\bhx \sim 0.5$. When observed at smaller IR wavelengths, the
cross-correlation actually traces the spectrum of the primordial stars
and strong features, such as Lyman-alpha peak could be clearly
reconstructed from the data.

We have roughly modeled the noise properties of several up-coming
experiments. Detection properties of experiments to detect
cross-correlations is often limited by a limited fraction of sky
observed with the first-generation of experiments. Nevertheless, we
predict that detection is within the reach of the next-generation
experiments. The most problematic at the moment is the IR field: the
early experiments such as CIBER or the AKARI NEP (North Ecliptic
Pole) survey are simply limited by a small fraction of the sky being
observed. A full-sky map at the sensitivity of AKARI NEP, will be
clearly able to detect the cross-correlation.  We note, that the
signal-to-noise ratios given in the Table \ref{tab2} are for a single
21-cm signal cross-correlation: stacking many of them will, of course,
improve the signal-to-noise ratio accordingly.

Finally, we note that our model is overly simplistic in many aspects.
Firstly, we assume a continuous Pop-III star-formation together with
continuous production of massive supernovae. In reality, however,
massive supernovae will pollute their environment with metals and
hence quench the Pop-III star formation, leading to a fast transition
to the Pop-II stars. Alternatively, supernovae are not so energetic
and Pop-III stars can actually completely reionize the Universe. In
the former case, our high-redshift calculations still hold, but at
lower redshifts both IR and SZ cross-correlations disappear as both
Pop-II star-light and the corresponding supernovae are much less
energetic.  In the latter case, IR correlates with 21-cm throughout
the duration of reionization, but we overestimated cross-correlation
with SZ. In either case, the cross-correlations can be used to
constrain reionization process.  Secondly, the constant bubble-size
used here cannot hold. In reality, the bubble size increases and gets
to the level of few Mpc at the peak of the reionization process, when
$x_H\sim0.5$ and the bubbles have not merged yet. Latter the bubbles
merge and the bubble-size is not a well-defined quantity. In this work
we used 4 $\mpch$ for a typical size of bubble and $2\mpch$ for the
typical size of sources.  Lowering the size of sources pushes the
auto-correlation and cross-correlation spectra up. However, the
relative amplitudes decreases and hence the prospects for detection
decrease.  Thirdly, the cross-correlation terms between the neutral
fraction and the density were ignored in this work, following the
analysis in \cite{Zaldarriaga:2003du}. While these terms might be
important, it is not immediately clear what would their effect be. An
analytical model for cross-correlation has been developed in
\cite{Furlanetto:2004nh} and the study of this effect is deferred to a
forthcoming publication.

In this paper we have put forward a novel idea for extracting
information of reionization.  Adding details to the model, such as
bubble size growth, neutral fraction -- density correlation, etc.,
either through a more accurate analytical modeling, but probably more
likely through insight from numerical codes, will probably improve and
somewhat change predictions presented here, but hopefully only at the
quantitative level.

\section*{Acknowledgments}
AS is supported by Oxford Astrophysics. AC thanks Peng Oh for useful discussions.

\label{lastpage}
\bibliography{../BibTeX/cosmo,../BibTeX/cosmo_preprints}
\bibliographystyle{mnras}
\bsp

\end{document}